\documentclass[aps,prb,twocolumn,showpacs,floatfix]{revtex4}

\usepackage{amsmath,amssymb,graphicx}

\begin{document}
\preprint{Submitted to Phys. Rev. B}


\title{Carrier dynamics and coherent acoustic phonons in nitride heterostructures}

\author{G. D. Sanders}
\author{C. J. Stanton}
\affiliation{Department of Physics, University of Florida, Box 118440\\
Gainesville, Florida 32611-8440}

\date{\today}


\begin{abstract}

We model generation and propagation of coherent acoustic phonons in piezoelectric
InGaN/GaN multi-quantum wells embedded in a \textit{pin} diode structure and compute
the time resolved reflectivity signal in simulated pump-probe experiments.
Carriers are created in the InGaN wells by ultrafast pumping below the GaN band
gap and the dynamics of the photoexcited carriers is treated in a Boltzmann equation
framework. Coherent acoustic phonons are generated in the quantum well via both
deformation potential electron-phonon and piezoelectric electron-phonon interaction
with photogenerated carriers, with the latter mechanism being the dominant one.
Coherent longitudinal acoustic phonons propagate into the structure at the sound
speed modifying the optical properties and giving rise to a giant oscillatory
differential reflectivity signal. We demonstrate that coherent optical control of the
differential reflectivity can be achieved using a delayed control pulse.

\end{abstract}



\maketitle

\section{Introduction}
\label{Introduction section}

Femtosecond transient reflectivity spectroscopy has proven useful in the study
of carrier dynamics in bulk semiconductors and semiconductor heterostructures as
well as the study of the generation and propagation of coherent phonons in a
number of materials. In particular, coherent optical phonons have been observed
in bulk semiconductors \cite{Dekorsy93.3842, Kuznetsov95.7555} and coherent
acoustic phonons have been detected in InGaN/GaN-based semiconductor heterostructures.
\cite{Chern04.339,Yahng02.4723,Stanton03.525,Liu05.195335,Sun01.1201,Ozgur01.5604}

We recently developed a detailed theory for carrier dynamics and time resolved
differential reflectivity in two-color pump-probe experiments on nonpiezoelectric
In$_x$Mn$_{1-x}$As/GaSb heterostructures, which agreed well with experimental
measurements.\cite{Sanders05.245302,Wang05.153311} Oscillations were observed
in the differential reflectivity which we attributed to the generation and
propagation of coherent acoustic phonon wave packets, which altered the local
dielectric function as they propagated through the structure at the LA phonon
sound speed. The carrier dynamics in the two-color pump-probe
experiments were modeled using a Boltzmann equation formalism including
photogeneration of carriers by the pump laser and their subsequent cooling
and relaxation by emission of confined LO phonons. The recombination of electron-hole
pairs via the Shockley-Read carrier trapping mechanism was included in a simple
relaxation time approximation.

The differential reflectivity was obtained by solving Maxwell's equations
throughout the structure using a transfer matrix method. In addition to the
coherent phonon induced reflectivity oscillations, a strong background signal was
observed which we attributed to: (1) enhanced Drude absorption by photoexcited
carriers, (2) relaxation dynamics due to cooling of hot photoexcited carriers
by LO phonon scattering, and (3) nonradiative recombination of electron-hole
pairs at midgap defects.

A microscopic theory for the generation and propagation of coherent acoustic phonons
in piezoelectic wurtzite semiconductors was developed in Ref.~\onlinecite{Sanders01.235316}
(see the erratum in Ref.~\onlinecite{Sanders02.079903} and the review article by
Chern \textit{et al} in Ref.~\onlinecite{Chern04.339}). It was found that generation of
coherent acoustic phonons by means of photoexcited carriers occurs through the
deformation potential electron-phonon interaction and the piezoelectric
electron-phonon interaction, with the latter typically being an order of
magnitude stronger. While the work of Ref.~\onlinecite{Sanders01.235316}
concerned itself with the generation and propagation of coherent acoustic
phonons in piezoelectric heterostructures, it did not address the question of
their detection in time resolved differential reflectivity experiments.

Our earlier paper (Ref.~\onlinecite{Sanders05.245302}) described the generation and
propagation of coherent acoustic phonons in Zinc Blende semiconductor heterostructures
in which the generation of coherent acoustic phonons can only occur by means of the
weaker deformation potential electron-phonon mechanism. In this paper we are
motivated to extend the results of Ref.~\onlinecite{Sanders05.245302} to piezoelectric
wurtzite heterostructures where the coherent phonon induced differential reflectivity
oscillations are expected to be much stronger.

\section{Theory}
\label{Theory section}

\begin{figure} [tbp]
\includegraphics[scale=.34]{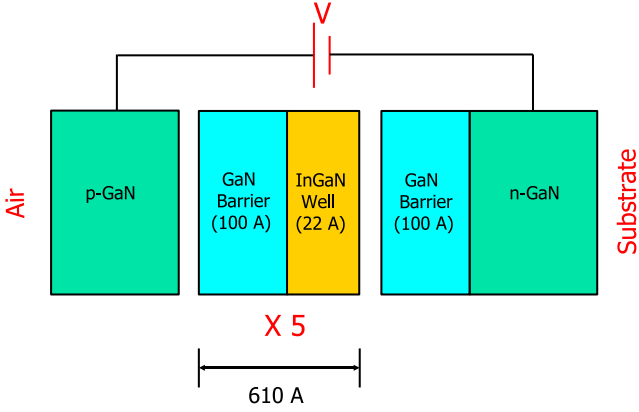}
\caption{
Schematic diagram of the multi-quantum well \textit{pin} diode structure.
}
\label{MQW diode figure}
\end{figure}

We model photogeneration of electrons and holes and the subsequent
excitation of coherent acoustic phonons in a nitride based multi-quantum well (MQW)
\textit{pin} diode shown schematically in Fig.~\ref{MQW diode figure}.
Such a nitride structures have been used in recent experimental studies by
Y. D. Jho \textit{et al} in Refs.~\onlinecite{Jho01.1130} and
\onlinecite{Jho02.035334}. The structures are grown on an undoped 1.5 $\mu$m thick
GaN layer grown on top of an $a_c$-plane sapphire substrate and we base our theoretical
studies on a similar structure. The intrinsic region of the diode is an undoped
MQW heterostructure consisting of five 22 {\AA} In$_{0.15}$Ga$_{0.85}$N wells
separated by 100 {\AA} GaN barriers. The growth direction is along (0001).
The multi-quantum well structure is the intrinsic region of a \textit{pin}
diode with contacts consisting of p- and n-doped GaN substrate and cap layers.
The thickness of the n-GaN cap layer is taken to be $100 \ \AA$, which is
thinner than that used in Refs.~\onlinecite{Jho01.1130} and \onlinecite{Jho02.035334},
and the width of the intrinsic region $L$ is 710 {\AA} as can be seen in
Fig.~\ref{MQW diode figure}. The voltage drop across the intrinsic region
is $\Delta V$. For the structure described in Refs.~\onlinecite{Jho01.1130}
and \onlinecite{Jho02.035334} the voltage drop across the intrinsic region
is $\Delta V \approx 1.6 \ \textrm{volts}$ when the external voltage across the
diode is $V = 0$ volts and we adopt this value in our studies.

\subsection{Bulk Bandstructure}

Our first task is to model the electronic states in the device shown in
Fig.~\ref{MQW diode figure}.
In bulk systems, the conduction and valence bands in wurzite crystals including
the effects of strain are treated using effective mass theory. Near the band
edge, the effective mass Hamiltonian for electrons is described by a $2 \times 2$
matrix which depends explicitly on electron wavevector, {\bf k}, and the strain
tensor, {\bf $\epsilon$}. The electron Bloch basis states are
\begin{subequations}
\label{conduction bloch state equation}
\begin{eqnarray}
\arrowvert c,1 \rangle &=& \arrowvert S \uparrow \rangle
\\
\arrowvert c,2 \rangle &=& \arrowvert S \downarrow \rangle.
\end{eqnarray}
\end{subequations}
The conduction band effective mass Hamiltonian is diagonal and
is given by (relative to the bottom of the conduction band)
(Refs.~\onlinecite{Jeon97.386} and \onlinecite{Chuang96.2491})
\begin{eqnarray}
H^{c}_{2 \times 2}({\bf k}, {\bf \epsilon}) = \{
\frac{\hbar^{2} k_{z}^{2}}{2 m^{*}_{z}} +
\frac{\hbar^{2} k_{t}^{2}}{2 m^{*}_{xy}} \nonumber \\
+ \ a_{c,z} \epsilon_{zz} + a_{c,xy} (\epsilon_{xx} + \epsilon_{yy}) \}
\ {\bf I}_{2 \times 2} .
\label{conduction band effective mass hamiltonian}
\end{eqnarray}
where ${\bf I}_{2 \times 2}$ is the identity matrix.
The electron effective masses along z (taken parallel to the c-axis) and in
the $xy$ plane are $m^*_z$ and $m^*_{xy}$, respectively, $k_t^2=k_x^2+k_y^2$,
and $\epsilon_{xx}$, $\epsilon_{yy}$ and $\epsilon_{zz}$ are strain tensor
components, and $a_{c,z}$ and $a_{c,xy}$ are the deformation potentials.

The Hamiltonian for the valence bands is a $6 \times 6$ matrix.
Following Ref. \onlinecite{Chuang96.1657}, the Hamiltonian (relative to the
top of the valence band) can be block diagonalized into two degenerate
$3 \times 3$ submatrices if we adopt the Bloch basis states
\begin{subequations}
\label{valence bloch state equation}
\begin{eqnarray}
\arrowvert v,1 \rangle &=&
- \frac{\alpha^{*}}{\sqrt{2}} \arrowvert (X + i Y) \uparrow \rangle
+ \frac{\alpha}{\sqrt{2}} \arrowvert (X - i Y) \downarrow \rangle
\\
\arrowvert v,2 \rangle &=&
\frac{\beta}{\sqrt{2}} \arrowvert (X - i Y) \uparrow \rangle
- \frac{\beta^{*}}{\sqrt{2}} \arrowvert (X + i Y) \downarrow \rangle
\\
\arrowvert v,3 \rangle &=&
\beta^{*} \arrowvert Z \uparrow \rangle
+ \beta \arrowvert Z \downarrow \rangle
\\
\arrowvert v,4 \rangle &=&
- \frac{\alpha^{*}}{\sqrt{2}} \arrowvert (X + i Y) \uparrow \rangle
- \frac{\alpha}{\sqrt{2}} \arrowvert (X - i Y) \downarrow \rangle
\\
\arrowvert v,5 \rangle &=&
\frac{\beta}{\sqrt{2}} \arrowvert (X - i Y) \uparrow \rangle
+ \frac{\beta^{*}}{\sqrt{2}} \arrowvert (X + i Y) \downarrow \rangle
\\
\arrowvert v,6 \rangle &=&
- \beta^{*} \arrowvert Z \uparrow \rangle
+ \beta \arrowvert Z \downarrow \rangle .
\end{eqnarray}
\end{subequations}
The phase factors, $\alpha$ and $\beta$, are functions of the angle
$\phi = \tan^{-1} (k_{y}/k_{x})$ and are given by
\begin{subequations}
\label{phases equation}
\begin{eqnarray}
\alpha(\phi) &=& \frac{1}{\sqrt{2}} \ e^{i(3 \pi /4 + 3 \phi /2) }
\\
\beta(\phi)  &=& \frac{1}{\sqrt{2}} \ e^{i(  \pi /4 +   \phi /2) } .
\end{eqnarray}
\end{subequations}
The block diagonalized Hamiltonian can be written as
\begin{equation}
H^{v}_{6 \times 6}({\bf k}, {\bf \epsilon}) =
\left(
\begin{array}{cc}
H^{U}_{3 \times 3}( {\bf k}, {\bf \epsilon} )  & 0 \\
0  & H^{L}_{3 \times 3}( {\bf k}, {\bf \epsilon} )
\end{array}
\right) ,
\label{valence band effective mass equation}
\end{equation}
where the upper and lower blocks of the Hamiltonian are
\begin{equation}
H^{U}_{3 \times 3}( {\bf k}, {\bf \epsilon} ) =
\left(
\begin{array}{ccc}
F      & K_t         & -iH_t         \\
K_t    & G           & \Delta-iH_t   \\
iH_t   & \Delta+iH_t & \lambda
\end{array}
\right)
\end{equation}
and
\begin{equation}
H^{L}_{3 \times 3}( {\bf k}, {\bf \epsilon} ) =
\left(
\begin{array}{ccc}
F      & K_t         &  iH_t         \\
K_t    & G           & \Delta+iH_t   \\
-iH_t  & \Delta-iH_t & \lambda
\end{array}
\right).
\end{equation}

The elements appearing in the $3 \times 3$ Hamiltonian matrices are
\begin{equation}
F = \Delta_1 + \Delta_2 + \lambda + \theta
\end{equation}
\begin{equation}
G = \Delta_1 - \Delta_2 + \lambda + \theta
\end{equation}
\begin{equation}
K_t = \frac{\hbar^2}{2 m_0} A_5 k_t^2
\end{equation}
\begin{equation}
H_t = \frac{\hbar^2}{2 m_0} A_6 k_t k_z
\end{equation}
\begin{equation}
\Delta = \sqrt{2} \ \Delta_3
\end{equation}
\begin{equation}
\lambda = \frac{\hbar^2}{2 m_0}\left( A_1 k_z^2 +A_2 k_t^2 \right)
+D_1 \epsilon_{zz}+ D_2 \left( \epsilon_{xx}+\epsilon_{yy} \right)
\end{equation}
\begin{equation}
\theta = \frac{\hbar^2}{2 m_0}\left( A_3 k_z^2 +A_4 k_t^2 \right)
+D_3 \epsilon_{zz}+ D_4 \left( \epsilon_{xx}+\epsilon_{yy} \right)
\end{equation}
Here, $m_0$ is the free electron mass, the $A_i's$ are effective mass parameters,
the $D_i's$ are the Bir-Pikus deformation potentials, and the $\Delta 's$ are
related to the crystal field splitting, $\Delta_{cr}$, and spin-orbit
splitting, $\Delta_{so}$, by $\Delta_1=\Delta_{cr}$ and
$\Delta_2=\Delta_3=\Delta_{so}/3$.

\subsection{Material parameters}

\begin{table}
\caption{Material parameters for wurtzite GaN and InN. Values are taken
from Ref.~\onlinecite{Vurgaftman01.5815} unless indicated otherwise.}
\begin{ruledtabular}
\begin{tabular} {lcr}
Parameter & GaN & InN \\
\tableline
\underline{Mass Density} & & \\
$\rho_0$   (gm/cm$^3$)    & 6.15    & 6.81 \\
\underline{Lattice constants}  & & \\
$a_0$( \AA )              & 3.189   & 3.545 \\
$c_0$( \AA )              & 5.185   & 5.703 \\
\underline{Direct band gaps}  & & \\
$E_g$       (eV)          & 3.507   & 1.994 \\
$\alpha_g$  (meV/K)       & 0.909   & 0.245 \\
$\beta_g$   (K)           & 830     & 624   \\
\underline{Hole splitting energies} & & \\
$\Delta_{cr}$ (eV)        & 0.019   & 0.041 \\
$\Delta_{so}$ (eV)        & 0.014   & 0.001 \\
\underline{Electron effective masses} & & \\
$m^{*}_{z}$  (m$_0$)      & 0.2     & 0.12 \\
$m^{*}_{xy}$ (m$_0$)      & 0.2     & 0.12 \\
\underline{Hole effective mass parameters} & &  \\
$A_1$                     & -6.56  & -8.21 \\
$A_2$                     & -0.91  & -0.68 \\
$A_3$                     &  5.65  &  7.57 \\
$A_4$                     & -2.83  & -5.23 \\
$A_5$                     & -3.13  & -5.11 \\
$A_6$                     & -4.86  & -5.96 \\
\underline{Electron deformation potentials} & & \\
$a_{c,z}$   (eV)          &  -6.5  & -3.5  \\
$a_{c,xy}$  (eV)          & -11.8  & -3.5  \\
\underline{Hole deformation potentials} & & \\
$D_1$       (eV)          &  -3.0  & -3.0  \\
$D_2$       (eV)          &   3.6  &  3.6  \\
$D_3$       (eV)          &   8.82 &  8.82 \\
$D_4$       (eV)          &  -4.41 & -4.41 \\
\underline{Elastic stiffness constants} & & \\
$C_{11}$    (GPa)         &  390   & 223 \\
$C_{12}$    (GPa)         &  145   & 115 \\
$C_{13}$    (GPa)         &  106   & 92  \\
$C_{33}$    (GPa)         &  398   & 224 \\
\underline{Piezoelectric constants} & & \\
$e_{13}$    (C/m$^2$)     & -0.35  & -0.57 \\
$e_{33}$    (C/m$^2$)     &  1.27  &  0.97 \\
\underline{Dielectric constants} & & \\
$\varepsilon_{0}$      & 8.9 \tablenotemark[1] & 15.3 \tablenotemark[2] \\
$\varepsilon_{\infty}$ & 5.7 \tablenotemark[3] & 8.4 \tablenotemark[3] \\
\underline{Refractive index} & & \\
$n$               & 2.27 \tablenotemark[4] & 2.71 \tablenotemark[5] \\
\end{tabular}
\end{ruledtabular}
\label{material parameter table}

\tablenotetext[1] {Ref. \ \onlinecite{Doshi98.2784}.}
\tablenotetext[2] {Ref. \ \onlinecite{Martin96.2541}.}
\tablenotetext[3] {Ref. \ \onlinecite{Bernardini97.3958}.}
\tablenotetext[4] {Ref. \ \onlinecite{Yu97.3209}.}
\tablenotetext[5] {Ref. \ \onlinecite{Yang02.9803}.}

\end{table}

The material parameters for GaN and InN are shown in
Table~\ref{material parameter table}. Unless otherwise stated, the
parameters are taken from Ref.~\onlinecite{Vurgaftman01.5815}.

The temperature dependent direct band gaps are given by the
empirical Varshni formula \cite{Varshni67.149}
\begin{equation}
E_g(T) = E_g - \frac{\alpha_g \ T^2}{T+\beta_g}
\label{varshni band gap equation}
\end{equation}
where $E_g$ is the band gap at absolute zero, $T$ is the temperature and
$\alpha_g$ and $\beta_g$ are the Varshni parameters. These band gap parameters
are listed in Table~\ref{material parameter table} for wurzite GaN and InN.

For the In$_{x}$Ga$_{1-x}$N alloy the temperature and composition dependent
band gap is given by
\begin{equation}
E_g(x,T)=x E_{g,InN}+(1-x)E_{g,GaN} - b x(1-x)
\label{alloy band gap equation}
\end{equation}
where $E_{g,InN}$ and $E_{g,GaN}$ are the temperature dependent
band gaps in bulk InN and GaN as given in Eq.~\ref{varshni band gap equation}
and $b=3.0 \ \mbox{eV}$ is the band gap bowing parameter for In$_{x}$Ga$_{1-x}$N.
\cite{Vurgaftman01.5815} While the value of the GaN band gap is well known, there is
some uncertainty in the value of the InN band gap used in computing
the band gap in In$_x$Ga$_{1-x}$N. The InN band gap determined from
absorption measurements on InN samples range from 1.7 to 3.1 eV with most
values centered around 1.9 eV. \cite{Walukiewicz04.300} Based on optical
measurements on ultrapure InN samples, it has recently been suggested
\cite{Walukiewicz04.300,Davydov02.1,Wu02.3967,Matsuoka02.1246} that InN is a
narrow gap semiconductor with a band gap of 0.7 eV as opposed to 1.994 eV as
cited in Ref.~\onlinecite{Vurgaftman01.5815}. Nevertheless, for the small values
of the Indium concentration, $x$, that we use in this study, the computed band
gap for In$_x$Ga$_{1-x}$N should be rather insensitive to the assumed value of
the InN band gap and we adopt the InN band gap and band gap bowing parameter
for the In$_x$Ga$_{1-x}$N alloy cited in Ref.~\onlinecite{Vurgaftman01.5815}.
Since the band bowing parameter, $b$, for In$_x$Ga$_{1-x}$N based on an InN band gap
of 0.7 eV has not been redetermined, our use of the older parameters has
the virtue of including band bowing effects in our calculations for
small values of $x$.

We obtain electron effective masses for In$_{x}$Ga$_{1-x}$N by
linearly interpolating the reciprocals of the masses as a function
of $x$, i.e., the concentration dependent effective masses are taken to
be
\begin{subequations}
\label{alloy electron effective mass equation}
\begin{eqnarray}
\frac{1}{m^*_{xy}(x)} &=& x \left(\frac{1}{m^*_{xy}}\right)_{\text{InN}}
+ \ (1-x) \left(\frac{1}{m^*_{xy}}\right)_{\text{GaN}}
\\
\frac{1}{m^*_{z}(x)} &=& x \left(\frac{1}{m^*_{z}}\right)_{\text{InN}}
+ \ (1-x) \left(\frac{1}{m^*_{z}}\right)_{\text{GaN}}.
\end{eqnarray}
\end{subequations}
For all other material parameters in Table~\ref{material parameter table},
we use linear interpolation in the composition to obtain values for the alloy.

\subsection{Carrier states in multi-quantum well diode}

In computing the electronic states in the intrinsic region of the pin diode shown
in Fig.~\ref{MQW diode figure}, we take the barriers
in \emph{the ohmic contacts} to be infinite for simplicity. The quantum confinement potentials
for electrons and holes in the intrinsic region arise from bandgap discontinuities
between well and barrier regions and the strain-induced piezoelectric field.
Thus the confinement potential is
\begin{equation}
V_\alpha(z,t)=V_{\alpha,\text{gap}}(z)+V_{\text{piezo}}(z),
\label{confinement potential equation}
\end{equation}
where $\alpha=\{c,v\}$ distinguishes between electrons and holes.

If the position-dependent band gap in the MQW is $E_g(z)$ as
determined from Eq.~\ref{varshni band gap equation}, then the confinement
potentials for electrons and holes are given by
\begin{subequations}
\label{band gap cofinement potential equation}
\begin{eqnarray}
V_{c,\text{gap}}(z)&=& E_{g,\min}+Q_c \ \left(E_g(z) - E_{g,\min} \right)
\\
V_{v,\text{gap}}(z)&=& - (1-Q_c)\ \left( E_g(z) - E_{g,\min} \right)
\end{eqnarray}
\end{subequations}
where $E_{g,\min} = \min_z [E_g(z)]$ is the minimum of the position
dependent band gap and $Q_c=0.6$ is the conduction band offset.\cite{Nakamura}

The confinement potential due to the strain-induced piezoelectric field
satisfies
\begin{equation}
\frac{dV_{\text{piezo}}(z)}{dz} = |e| \ E^0_z(z),
\label{piezoelectric confimement potential equation}
\end{equation}
where $|e|$ is the electric charge and $E^0_z(z)$ is the
strain-induced piezoelectric field. The piezoelectric field in the
diode is obtained from the requirement that the electric displacement
vanishes.\cite{Smith88.2717} Thus
\begin{equation}
E_z^0(z)=-\ \frac{4 \pi}{\varepsilon_0(z)} \
\left( P_z^0(z) + P_0 \right),
\label{EzEq}
\end{equation}
where $P_0$ is a constant polarization induced by the externally applied
voltage, $\Delta V$, and $\varepsilon_0(z)$ is the position-dependent
static dielectric constant. The value of $P_0$ is obtained from the voltage
drop across the diode (of length $L$). The voltage drop in
Fig.~\ref{MQW diode figure} between source and drain due to the
induced piezoelectric field is just
\begin{equation}
\Delta V = - \int_{0}^{L} dz \ E_z^0(z) ,
\label{P0 equation}
\end{equation}
from which $P_0$ can be determined.
The magnitude of the strain-induced polarization directed along the
z-direction is given by
\begin{equation}
P_z^0(z)= e_{31}(z) \left( \epsilon_{xx}(z)+\epsilon_{yy}(z) \right)
+ e_{33}(z) \epsilon_{zz}(z),
\label{strain induced polarization equation}
\end{equation}
where $e_{31}(z)$ and $e_{33}(z)$ are the position-dependent piezoelectric
constants and $\epsilon_{xx}(z)$, $\epsilon_{yy}(z)$ and $\epsilon_{zz}(z)$
are the position-dependent strain tensor components. The orientation of
the strain-induced polarization, $P_z^0(z)$, is such that the piezoelectric
field opposes the built-in electric field in the intrinsic region as has been
discussed in several previous studies.
\cite{Jho01.1130, Lefebvre01.1252, Takeuchi98.1691, Im98.9435, Wetzel00.2159, Chichibu00.5153}

In a pseudomorphically strained MQW diode, the GaN source and drain contacts
are assumed to be unstrained while the in-plane lattice constants in the
In$_{x}$Ga$_{1-x}$N wells and GaN barriers adjust to the lattice constant in the
source and drain. For a MQW grown along the [0001] direction (the z-direction)
the z-dependent strain is given by \cite{Wright97.2833}
\begin{equation}
\epsilon_{xx}(z)=\epsilon_{yy}(z)=\frac{a_0-a(z)}{a(z)}.
\label{xx yy strain tensor equation}
\end{equation}
Here $a_0$ is the lattice constant in the GaN source and drain and $a(z)$ is
the z-dependent lattice constant in the MQW active region. Minimizing the
overall strain, we find \cite{Wright97.2833}
\begin{equation}
\epsilon_{zz}(z) = - \ \frac{2 \ C_{13}(z)}{C_{33}(z)} \ \epsilon_{xx}(z),
\label{zz strain tensor equation}
\end{equation}
where $C_{13}(z)$ and $C_{33}(z)$ are z-dependent elastic stiffness constants.

The diode structure breaks translational symmetry along the $z$ direction.
Thus, quantum confinement of carriers in the MQW active region gives rise
to a set of two-dimensional subbands. The effective mass wavefunctions are
\begin{equation}
\psi^{\alpha}_{n,{\bf k}}({\bf r})= \sum_{j}\
\frac{ e^{i \ {\bf k} \cdot {\bf \rho} } }{\sqrt{A}} \ F^{\alpha}_{n, k, j}(z)
\ \arrowvert \alpha, j \rangle ,
\label{Wavefunction equation}
\end{equation}
where $\alpha = \{ c,v \}$ refers to conduction or valence subbands,
$n$ is the subband index, ${\bf k} = (k_x,k_y,0) = (k,\phi)$ is the
two-dimensional wavevector, and $j$ labels the spinor component. For
conduction subbands, ($\alpha$=c)\ $j = 1,2$ \ while for valence subbands
($\alpha$=v)\ $j=1 ... 6$. The slowly varying envelope functions
$F^{\alpha}_{n, k, j}(z)$ are real and depend only on $k = |{\bf k}|$, while
the rapidly-varying Bloch basis states $\arrowvert \alpha, j \rangle$ are
defined in Eqs.~(\ref{conduction bloch state equation}) and
(\ref{valence bloch state equation}). The valence Bloch basis states depend on
the orientation, $\phi$, of $\textbf{k}$ in the $xy$ plane as given in
Eq.~(\ref{phases equation}). The area of the MQW sample in the $xy$ plane
is $A$, and ${\bf \rho} = (x,y,0)$ is the projection of {\bf r} in the plane.

The envelope functions satisfy a set of effective-mass Schr\"{o}dinger
equations
\begin{equation}
\sum_{j,j'}  \left\{ H^{\alpha}_{j,j'}( k )
+ \delta_{j,j'}\ \left[ V_{\alpha}(z)-E^{\alpha}_n( k ) \right] \right\}
F^{\alpha}_{n, k, j' }(z) =0,
\label{Schrodinger equation}
\end{equation}
subject to the boundary conditions
\begin{equation}
F^{\alpha}_{n, k, j }(z=0) \ = \ F^{\alpha}_{n, k, j }(z=L) \ = \ 0.
\label{Boundary condition equation}
\end{equation}
where $V_{\alpha}(z)$ are the quantum confinement potentials
for conduction and valence electrons defined in
Eq.~\ref{confinement potential equation} and $E^{\alpha}_n( k )$ are the energy
eigenvalues for the $n$th conduction or valence subband.  Note that in the
envelope function approximation, the subband energy depends only on the
magnitude $k$ of the transverse wavevector and not on the angle $\phi$.
The matrix operators $H^{\alpha}_{j,j'}(k)$ depend on $z$ and are obtained
by making the replacement $k_{z} \rightarrow -i \frac{\partial}{\partial z}$
and letting all material parameters be $z$-dependent operators in the matrices
$H^{\alpha}(k,\epsilon)$ given in
Eqs.~(\ref{conduction band effective mass hamiltonian})
and (\ref{valence band effective mass equation}).
To ensure the Hermitian property of the Hamiltonian, we make the
operator replacements \cite{Chuang}
\begin{equation}
B(z) \ \frac{\partial^2}{\partial z^2} \rightarrow
\frac{\partial}{\partial z} \ B(z) \ \frac{\partial}{\partial z},
\end{equation}
and
\begin{equation}
B(z) \ \frac{\partial}{\partial z} \rightarrow
\frac{1}{2} \ \left[ B(z) \ \frac{\partial}{\partial z}
+ \frac{\partial}{\partial z} \ B(z)  \right] .
\end{equation}

We arrive at a set of coupled ordinary differential equations (ODE's) subject
to the two-point boundary value condition of Eq.\ (\ref{Boundary condition equation}).
These are solved for the envelope functions and subband energies.
In practice, we introduce a uniform grid, $\{ z_i \}$, along the z-direction
and finite-difference the effective mass Schr\"{o}dinger equations
to obtain a matrix eigenvalue problem which can be solved using
standard matrix eigenvalue routines. The resulting eigenvalues are the subband
energies, $E^{\alpha}_n( k )$, and the corresponding eigenvectors are
the envelope functions, $F^{\alpha}_{n,k,j}(z_i)$, defined on the finite
difference mesh.

\subsection{Carrier dynamics}

In two-color time resolved differential reflectivity experiments
a pump laser is used to excite electrons from the valence to the
conduction subbands of the multi-quantum well. The photoexcited
carriers then relax through scattering, changing the optical properties
of the heterostructure in the process. We simulate these processes using Boltzmann
transport equations which we solve numerically.\cite{Sanders98.9148,Sanders05.245302}
For each electron and hole subband state with energy, $E^\alpha_n(k)$, we have
a time dependent distribution function, $f^\alpha_n(\mathbf{k},t)$,
which gives the probability, as a function of time, of finding a conduction
or valence electron in subband $n$ with wave vector $\mathbf{k}$.
The Boltzmann equation including photoexcitation of hot electron-hole
pairs by the pump, the subsequent cooling of these carriers by emission
of confined LO phonons, and the recombination of electron-hole pairs is
\begin{eqnarray}
\nonumber && \frac{\partial f^\alpha_n(\mathbf{k})}{\partial t}
= \sum_{n',k'} \{ f^\alpha_{n'}(\mathbf{k}')
\ W_{\mathbf{k}',\mathbf{k}}^{n',n}
\ \left[ 1-f^\alpha_n(\mathbf{k}) \right] \\
&& - f^\alpha_n(\mathbf{k})
\ W_{\mathbf{k},\mathbf{k}'}^{n,n'}
\ \left[ 1-f^\alpha_{n'}(\mathbf{k}') \right] \}
+ \left[ \frac{\partial f^\alpha_n(\mathbf{k})}{\partial t} \right]
\label{2D Boltzmann equation}
\end{eqnarray}

To simplify the calculations, we use an axial approximation in which
the distribution functions are replaced by their angular averages in
the $x$-$y$ plane of the multi-quantum well.\cite{Sanders98.9148,Sanders05.245302}
Thus the time dependent conduction and valence electron distribution functions,
$f^c_n(k)$ and $f^v_{n'}(k)$, are functions of the magnitude of $\mathbf{k}$.

The first two terms on the right hand side of Eq.~(\ref{2D Boltzmann equation})
describes the rapid cooling of hot electrons and holes by emission
and absorption of confined LO phonons. Rapid LO phonon scattering is an
intraband process which creates quasi-thermal electron and hole
distributions at their respective band edges.
The confined LO phonon scattering rate in the multi-quantum well,
$W_{\mathbf{k},\mathbf{k}'}^{n,n'}$,
given in Refs.~\onlinecite{Sanders98.9148} and \onlinecite{Sanders05.245302},
is the rate at which a conduction (or valence) electron in subband $n$
with wave vector $\mathbf{k}$ scatters to subband $n'$ with
wave vector $\mathbf{k}'$.

The last term on the right hand side of Eq.~(\ref{2D Boltzmann equation})
consists of two terms
\begin{equation}
\left[ \frac{\partial f^\alpha_n(\mathbf{k})}{\partial t} \right] =
\left[ \frac{\partial f^\alpha_n(\mathbf{k})}{\partial t} \right]_{pump} +
\left[ \frac{\partial f^\alpha_n(\mathbf{k})}{\partial t} \right]_{relax}
\label{pump and relaxation equation}
\end{equation}
and describes the change in the conduction or valence electron distribution
function due to the action of the pump as well as carrier relaxation through
recombination of electron-hole pairs. We assume that electron-hole pairs can
recombine with a phenomenological time constant $\tau_0$.
Expressions for the photogeneration rate can be found in
Ref.~\onlinecite{Sanders05.245302}. It is assumed that the pump laser
is a Gaussian pulse with an intensity FWHM of $\tau_p$ peaked at t = 0
with a Lorentzian spectral lineshape with a FWHM of $\gamma$.

\subsection{Optical properties}

For dipole transitions between conduction and valence electronic states near the
band edge, we calculate optical properties using Fermi's golden rule. The
contributions to the real and imaginary parts of the dielectric function from
the effective mass band edge states are given by
\begin{eqnarray}
\label{eps1 equation}
\varepsilon_1(\hbar\omega) &=& \frac{8 \pi e^2}{V}
\sum_{n,n',\textbf{k}}
\left| \hat{{\bf \epsilon}} \centerdot {\bf d}^{c,v}_{n,n'}({\bf k}) \right|^2
\left( f^v_{n'}(k) - f^c_{n}(k) \right)
\nonumber \\
&\times& \frac{\Delta E^{c,v}_{n,n'}(k)}
{\left(\Delta E^{c,v}_{n,n'}(k)+\hbar\omega\right)
\left(\Delta E^{c,v}_{n,n'}(k)-\hbar\omega\right)}
\end{eqnarray}
and
\begin{eqnarray}
\label{eps2 equation}
\varepsilon_2(\hbar\omega) &=& \frac{4 \pi^2 e^2}{V}
\sum_{n,n',\textbf{k}}
\left| \hat{{\bf \epsilon}} \centerdot {\bf d}^{c,v}_{n,n'}({\bf k}) \right|^2
\left( f^v_{n'}(k) - f^c_{n}(k) \right)
\nonumber \\ &\times&
\delta(\Delta E^{c,v}_{n,n'}(k)-\hbar\omega)
\end{eqnarray}
where $\Delta E^{c,v}_{n,n'}(k) = E^c_n(k)-E^v_{n'}(k)$ are the $k$-dependent
transition energies between conduction and valence subband states, $V = A \ L$
is the sample volume, and $\hat{{\bf \epsilon}}$ is the unit complex polarization
vector. Expressions for the dipole matrix elements ${\bf d}^{c,v}_{n,n'}({\bf k})$
for dipole allowed transitions between conduction subband  $n$ and and valence
subband $n'$ can be found in Ref.~\onlinecite{Sanders01.235316}.

In Eqs.~(\ref{eps1 equation}) and (\ref{eps2 equation}) the
spectral lineshapes are described by Dirac delta functions.
In practice we replace the delta functions with
Lorentzian lineshapes having a full width at half maximum
(FWHM) of $\Gamma$. The real and imaginary parts
of the dielectric function, including lineshape broadening,
can be obtained from Eqs.~(\ref{eps1 equation}) and
(\ref{eps2 equation}) using the operator replacements
\cite{Chuang}
\begin{equation}
\delta(\Delta E-\hbar\omega) \rightarrow
\frac{\Gamma /(2\pi)} {{(\Delta E-\hbar\omega)^2+(\Gamma/2)^2}}
\label{eps2 broadening equation}
\end{equation}
and
\begin{equation}
\frac{1}{\Delta E-\hbar\omega} \rightarrow
\frac{\Delta E-\hbar\omega}{(\Delta E-\hbar\omega)^2+(\Gamma/2)^2}.
\label{eps1 braodening equation}
\end{equation}

\subsection{Generation and propagation of coherent acoustic phonons}

The ultrafast photogeneration of electrons and holes in the InGaN
multi-quantum wells by the pump gives rise to coherent longitudinal
acoustic (LA) phonons which propagate into the diode structure.
Coherent acoustic phonons give rise to a macroscopic lattice displacement.
\cite{Sanders01.235316,Sanders02.079903,Chern04.339,Stanton03.525,Yahng02.4723,Liu05.195335,Sun01.1201,Ozgur01.5604}
Since the photogenerated carrier distributions are functions of $z$, the
transient lattice displacement $U(z,t)$ due to photogenerated carriers
is independent of $x$ and $y$ and is parallel to $z$. As discussed in
Refs. \onlinecite{Sanders01.235316} and \onlinecite{Chern04.339}, $U(z,t)$
satisfies a loaded string equation. In the presence of a position
dependent longitudinal acoustic sound velocity, $C_s(z)$, we have
\begin{equation}
\frac{\partial^2 U(z,t)}{\partial t^2} -
\frac{\partial}{\partial z}
\left( C^2_s(z) \ \frac{\partial U(z,t)}{\partial z}
\right)
=S(z,t)
\label{Loaded string equation}
\end{equation}
where $S(z,t)$ is a driving or loading function and depends on
the photogenerated carrier distributions.
The longitudinal acoustic sound velocity is given by
\begin{equation}
C_s(z) = \sqrt{\frac{C_{33}(z)}{\rho_0(z)}}
\label{Sound velocity}
\end{equation}
where $C_{33}(z)$ and $\rho_0(z)$ are the position dependent
elastic stiffness constant and mass density.

The loaded string equation is solved subject to the initial conditions
\begin{equation}
U(z,t=-\infty) = \frac{\partial U(z,t=-\infty)}{\partial t} = 0.
\label{U(z,t) initial conditions}
\end{equation}
We solve the loaded string equation numerically by finite
differencing Eq. (\ref{Loaded string equation}) inside a
computational box whose left edge, $z_L$, is the semiconductor-air
interface in the p-GaN contact and whose right edge, $z_R$, lies
inside the n-GaN substrate (see Fig. \ref{MQW diode figure}).
At $z_R$ we impose absorbing boundary conditions while at $z_L$
there are no perpendicular forces at the semiconductor-air interface.
Thus we solve the initial value problem subject to the left
and right boundary conditions
\begin{subequations}
\label{U(z,t) boundary conditions}
\begin{equation}
\frac{\partial U(z_L,t)}{\partial z} = 0
\end{equation}
and
\begin{equation}
\frac{\partial U(z_R,t)}{\partial z}+
\frac{1}{C_s(z_R)} \ \frac{\partial U(z_R,t)}{\partial t}=0.
\end{equation}
\end{subequations}

Starting with the second quantized Hamiltonian for the electron-phonon
interaction, a microscopic expression for the driving function was
derived in Ref. \onlinecite{Sanders01.235316} using the density
matrix formalism (see also the erratum in
Ref. \onlinecite{Sanders02.079903} as well as the review
article in Ref. \onlinecite{Chern04.339}).
In wurzite materials such as InGaN the electron-phonon interaction
contains contributions from piezoelectric and deformation potential
coupling. Under typical experimental conditions the microscopic
expression for the driving function can be simplified
to \cite{Sanders01.235316,Sanders02.079903}
\begin{equation}
S(z,t)=\sum_\nu S_\nu(z,t)
\label{Szt sum}
\end{equation}
where the summation index, $\nu$, runs over carrier species, i.e.
electrons, heavy holes, light holes and split off holes. The
partial driving functions including both piezoelectric and deformation
potential coupling are given by
\begin{equation}
S_\nu(z,t) = \pm \frac{1}{\rho_0}
\left(
a_\nu \frac{\partial}{\partial z}+\frac{4 \pi |e| \ e_{33}}{\epsilon_\infty}
\right) \rho_\nu(z,t)
\label{Simplified S(z,t)}
\end{equation}
where $\rho_0$ is the mass density, $e_{33}$ is the piezoelectric constant,
$\epsilon_\infty$ is the high frequency dielectric constant and $\rho_\nu(z,t)$
is the photogenerated carrier density. The deformation potentials are
$a_\nu$. For conduction electrons, $a_\nu = a_{c,z}$, for heavy and light
holes, $a_\nu = D_1 + D_3$, and for crystal field split holes, $a_\nu = D_1$.
We note that Eq.~(\ref{Simplified S(z,t)}) was derived in the elastic continuum
limit by Chern \textit{et al}. \cite{Chern04.339}
The driving function satisfies the sum rule
\begin{equation}
\int_{-\infty}^{\infty} dz \ S(z,t) = 0.
\label{Sum rule}
\end{equation}
as shown in Refs. \onlinecite{Sanders01.235316} and
\onlinecite{Chern04.339}.

Since the photoexcited
holes are predominantly a mixture of heavy and light holes, we use
$a_{\nu}=D_{13}=D_1+D_3$ in computing hole deformation potential contributions in
Eq.~(\ref{Szt sum}). The sum over species, $\nu$, then yields the total
driving function
\begin{eqnarray}
S(z,t)&&= \frac{1}{\rho_0} \left\{
a_{c,z} \frac{\partial}{\partial z}+\frac{4 \pi |e| e_{33}}{\epsilon_{\infty}}
\right\} \rho_{\text{e}}(z,t)
\nonumber \\
&& - \ \frac{1}{\rho_0} \left\{
D_{13} \frac{\partial}{\partial z}+\frac{4 \pi |e| e_{33}}{\epsilon_{\infty}}
\right\} \rho_{\text{h}}(z,t) ,
\label{Snu2}
\end{eqnarray}
where $\rho_{\text{e}}(z,t)$ and $\rho_{\text{h}}(z,t)$ are
the photogenerated conduction electron and valence hole densities.

The photogenerated electron and hole densities are obtained from
the envelope functions defined in Eq.~(\ref{Wavefunction equation})
and the time dependent electron and hole distribution
functions, $f^c_n(k,t)$ and $f^v_{n'}(k,t)$, obtained from solving
the coupled Boltzmann equations (\ref{2D Boltzmann equation}).
We have
\begin{subequations}
\label{Photogenerated carrier densities}
\begin{equation}
\rho_e(z,t)=\frac{1}{A} \sum_{n,j,\mathbf{k}}
\left( f^c_n(k,t)-f_n^{c0}(k) \right)
\left| F^c_{n,k,j}(z) \right|^2
\end{equation}
and
\begin{equation}
\rho_h(z,t)=\frac{1}{A} \sum_{n',j,\mathbf{k}}
\left( f_{n'}^{v0}(k)-f^v_{n'}(k,t) \right)
\left| F^v_{n',k,j}(z) \right|^2.
\end{equation}
\end{subequations}
where $f_n^{c0}(k)$ and $f_{n'}^{v0}(k)$ are the initial Fermi-Dirac
distribution functions for conduction and valence band electrons
in subbands $n$ and $n'$.

The propagating coherent phonon displacement field $U(z,t)$ gives
rise to a propagating strain with
\begin{equation}
\varepsilon_{zz}(z,t)=\frac{\partial U(z,t)}{\partial z}
\label{Strain tensor components}
\end{equation}
The propagating strain field alters the optical properties of the
sample which is then detected by the delayed probe pulse.

\subsection{Transient probe response}

\begin{figure} [tbp]
\includegraphics[scale=.75]{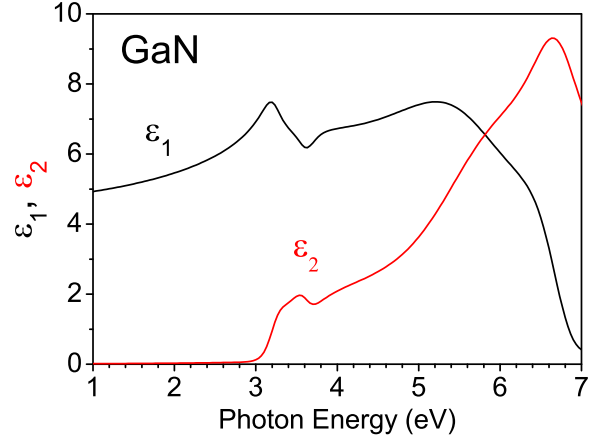}
\caption{(Color Online)
Model dielectric function for bulk GaN at $T = 0 \ \mbox{K}$ as a function of
photon energy used in calculating the dielectric function in the diode source
and drain contact regions.
}
\label{GaN dielectric function figure}
\end{figure}

To compute the time dependent probe transmission and reflection
coefficients we need to model the dielectric function in the diode
structure shown in Fig.~\ref{MQW diode figure}. We will denote this
time and position dependent dielectric function as
\begin{equation}
\varepsilon(\hbar\omega,z,t) =
\varepsilon_1(\hbar\omega,z,t) + i \ \varepsilon_2(\hbar\omega,z,t)
\label{Total dielectric function}
\end{equation}
where $\hbar \omega$ is the probe energy. We can solve Maxwell's equations
for the time dependent probe reflection coefficient, $R(\hbar\omega,t)$,
using the transfer matrix method of Ref.~\onlinecite{Sanders05.245302}
as described in Ref. \onlinecite{Chuang}.

There are several processes which contribute to the dielectric function
in the InGaN/GaN multi-quantum wells. The first contribution to the
dielectric function is a Drude term due to photoexcited free carriers
which gives a contribution to the dielectric function of
\begin{equation}
\varepsilon(\hbar\omega,z,t)_{D}=
-\frac{\left(\hbar\omega_p(t) \right)^2}{\left( \hbar\omega \right)^2}
\label{Drude term}
\end{equation}
where $\omega_p$ is the plasma frequency. The Drude contribution
to the dielectric function in Eq.~(\ref{Drude term}) is uniform
in the multi-quantum well structure and vanishes everywhere else.
In the random phase approximation (RPA), the time dependent plasma
frequency is given by \cite{Schafer}
\begin{equation}
\omega_p^2(t)=\frac{4 \pi e^2}{LA}
\left[
\sum_{n,\mathbf{k}}
\frac{f^c_n(k,t)}{m_n^*(k)}+
\sum_{n',\mathbf{k}}
\frac{f^v_{n'}(k,t)-1}{m_{n'}^*(k)}
\right]
\label{Plasma frequency}
\end{equation}
where $L = 710 \ \AA$ is the width of the multi-quantum well structure
in Fig.~\ref{MQW diode figure} and $A$ is the cross sectional area of the
diode. The effective masses for electrons and holes are obtained from
the computed energy bands as
\begin{equation}
\frac{1}{m_n^*(k)}=\frac{1}{\hbar^2} \
\frac{\partial^2 E^\alpha_n(k)}{\partial k^2}.
\label{Plasma frequency effective mass}
\end{equation}
Note, that in our convention, the bandstructure is computed in the electron
picture so that the hole masses in Eq.~(\ref{Plasma frequency effective mass})
are usually negative. Thus both terms in Eq.~(\ref{Plasma frequency}) give
positive contributions to the squared plasma frequency.

Another contribution to the multi-quantum well dielectric function,
which we denote $\varepsilon(\hbar\omega,z,t)_{MQW}$, comes from
interband transition between the effective mass band edge states. The real and imaginary
parts of $\varepsilon(\hbar\omega,z,t)_{MQW}$ are given in Eqs.~(\ref{eps1 equation})
and (\ref{eps2 equation}) and are time dependent due to variations in the
electron and hole distribution functions with time.

There is also a background dielectric function, $\varepsilon_b(\hbar\omega)$
due to all the higher lying electronic transitions. For simplicity,
we treat these contributions to the dielectric function in the multi-quantum
well region using the model dielectric function for bulk GaN in
Ref.~\onlinecite{Djurisic99.2848} with contributions from the
$E_0$ and $E_0+\Delta_0$ critical points removed. These correspond
to contributions from the multi-quantum well effective mass electronic
states and are already included in Eqs.~(\ref{eps1 equation})
and (\ref{eps2 equation}).

We assume the coherent phonon alter the background dielectric function because
of strain induced variations in the energy gaps associated with each transition.
\cite{Thomsen84.989,Thomsen86.4129}
Density functional calculations of the deformation potentials
for the $E_1$ transitions in a number of semiconductors \cite{Ronnow99.5575}
have shown that the deformation potentials associated with the $E_0$ and $E_1$
features are equal to within 20\%. Thus, at photon energies at or below the
$E_0$ and $E_1$ transitions we assume that the effect of temperature and strain
on $\varepsilon_b$ is to introduce a \emph{rigid shift} in the dielectric function
such that
\begin{equation}
\varepsilon_{b}(\hbar\omega,z,t)=
\varepsilon_{b}(\hbar\omega-\Delta E_g(T)-
a_{cv}(\varepsilon_{xx}+\varepsilon_{yy}+\varepsilon_{zz})).
\label{Background dielectric function shift}
\end{equation}
Here, $a_{cv}$ is an effective deformation potential which we
take to be $a_{cv} = -18.3 \ \mbox{eV}$.
The coherent phonon strain tensor components are, $\varepsilon_{xx}$, $\varepsilon_{yy}$,
and $\varepsilon_{zz}$, and $\Delta E_g(T) = E_g(T)-E_g$ is the band gap shift due to
temperature variations with $E_g(T)$ being the temperature dependent band
gap defined by the empirical Varshni formula in Eq.~(\ref{varshni band gap equation}).

The total dielectric function in the multi-quantum well region is
obtained by adding the Drude, multi-quantum well, and background contributions in
Eqs.(\ref{Drude term}), (\ref{eps1 equation}), (\ref{eps2 equation}) and
(\ref{Background dielectric function shift}), i.e.
\begin{equation}
\varepsilon(\hbar\omega,z,t)=
\varepsilon(\hbar\omega,z,t)_{D}
+\varepsilon(\hbar\omega,z,t)_{MQW}
+\varepsilon_{b}(\hbar\omega,z,t)
\label{Sum total of dielectric functions}
\end{equation}

In the GaN contact regions, we use the model dielectric functions
for bulk GaN described in Ref. \onlinecite{Djurisic99.2848}.
Fig.~\ref{GaN dielectric function figure} shows the real and imaginary
parts of the model dielectric function for bulk GaN at
$T = 0 \ \mbox{K}$ in the absence of strain. Temperature and
strain effects are included using the same rigid shift model
as defined in Eq. (\ref{Background dielectric function shift}).
Note that the dielectric functions in the GaN source and drain
regions are modulated by the coherent phonon strain field as it
propagates through the structure.

\section{Results}
\label{Results section}

\subsection{Multi-quantum well band structure and electronic states}

\begin{figure} [tbp]
\includegraphics[scale=.75]{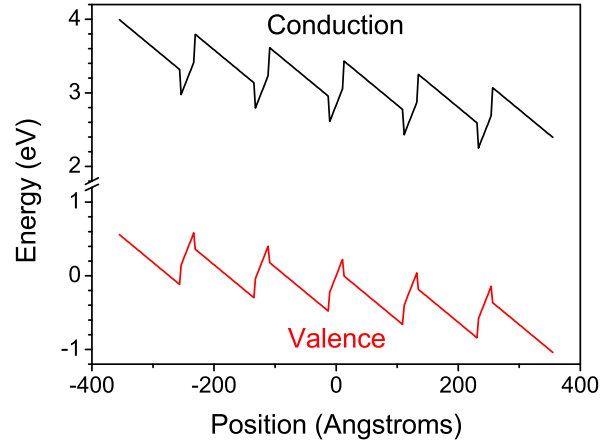}
\caption{(Color Online)
Band diagram for conduction and valence bands in the intrinsic region
of an InGaN/GaN MQW \textit{pin} diode structure. For an external
bias of $V \approx 0 \ \textrm{Volts}$, the voltage drop across the intrinsic region
of the diode is assumed to be
$\Delta V \approx 1.6 \ \textrm{Volts}$ as described in the text.
}
\label{confinement potential figure}
\end{figure}
%
The band diagram for the conduction and valence subband states in the intrinsic
region of our pseudomorphically strained InGaN/GaN MQW \textit{pin} diode
structure, shown schematically in Fig.~\ref{MQW diode figure}, can be seen in
Fig.~\ref{confinement potential figure} as a function of position along the
growth direction, $z$.
The voltage drop across the intrinsic region of the
diode is assumed to be $\Delta V \approx 1.6 \ \textrm{Volts}$ which is the
experimentally measured value obtained in similar structures described in
Refs.~\onlinecite{Jho01.1130} and \onlinecite{Jho02.035334} for an externally
applied voltage of $V \approx  0  \ \textrm{Volts}$. The center of the MQW
intrinsic region is defined to be at $z = 0$ and the valence band potential is
set to zero at this point. Since the in-plane lattice constant throughout the
structure is the same as in the GaN source and drain contacts, the GaN barriers
in the MQW region are unstrained while the In$_{0.15}$Ga$_{0.85}$N wells are
under biaxial compression. The electric field is 2.118 Mv/cm in the InGaN wells
and -0.686 Mv/cm in the GaN barriers.

\begin{figure} [tbp]
\includegraphics[scale=.9]{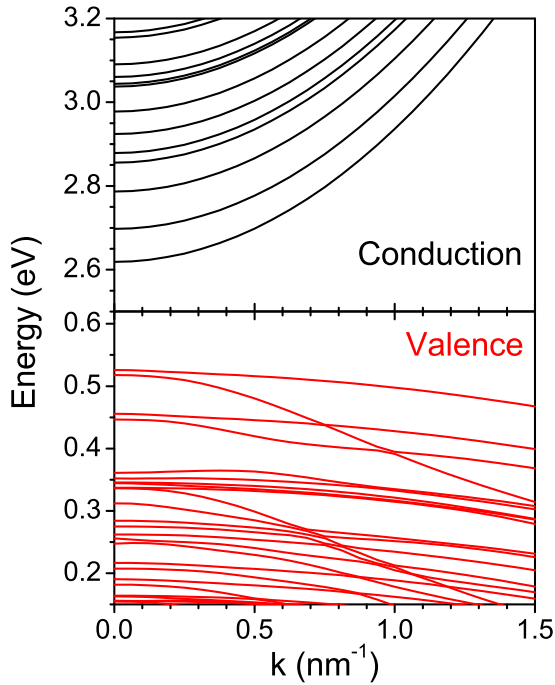}
\caption{(Color Online)
Conduction and valence subbands for the InGaN/GaN MQW \textit{pin} diode
structure of Fig.~\ref{MQW diode figure}
with an external bias of $V \approx 0 \ \textrm{Volts}$.
}
\label{En(k) figure}
\end{figure}

Using the confinement potentials in Fig.~\ref{confinement potential figure}
we can solve the effective mass Schr\"{o}dinger equation (\ref{Schrodinger equation})
for the electronic subbands $E^{\alpha}_n( k )$ and the envelope functions,
$F^{\alpha}_{n, k, j }(z)$. We can then compute the optical dipole matrix
elements ${\bf d}^{c,v}_{n,n'}({\bf k})$.
The conduction and valence subbands, $E^{\alpha}_n( k )$, for the
confinement potentials in Fig.~\ref{confinement potential figure} are shown
in Fig.~\ref{En(k) figure} as functions of $k$ in the axial approximation.
Note that the scale is different for the conduction and valance band levels
in the upper and lower panels.
%

\subsection{Multi-quantum well differential absorption}

Given the conduction and valence band distribution functions $f^c_{n}(k)$ and
$f^v_{n}(k)$, the real and imaginary parts of the dielectric function as a function
of photon energy can be obtained from Eqs.~\ref{eps1 equation} and \ref{eps2 equation}.
%
\begin{figure} [tbp]
\includegraphics[scale=.75]{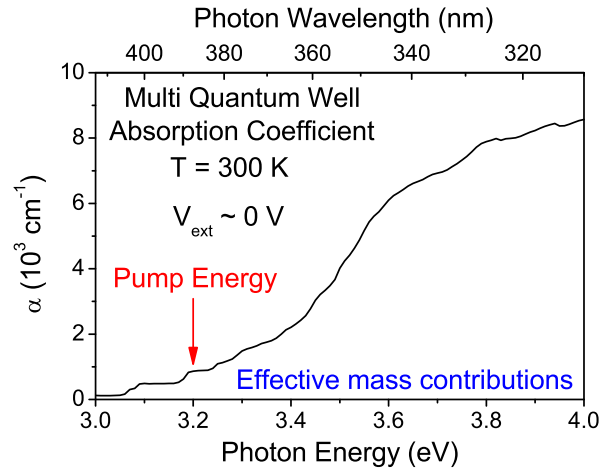}
\caption{(Color Online)
Initial absorption coefficient at room temperature due to optical transitions
between band edge effective mass states in an InGaN/GaN MQW \textit{pin} diode
structure with an external bias of $V \approx 0 \ \textrm{Volts}$.
}
\label{initial absorption coefficient figure}
\end{figure}
%
The room temperature absorption coefficient in the \textit{pin} diode
MQW intrinsic region due to intersubband optical transitions between the
effective mass states is shown in Fig.~\ref{initial absorption coefficient figure}.

To illustrate the effects of carrier dynamics on the optical properties,
we simulate a two-color pump-probe experiment on the diode structure with a pump
laser having a photon energy of 3.2 eV ($\lambda = 387.5 \ \textrm{nm}$) as
indicated in Fig.~\ref{initial absorption coefficient figure}.
The pump fluence is assumed to be .001 J/cm$^2$ and the FWHM temporal and
spectral widths are taken to be $\tau_p = 1 \ \mbox{ps}$ and $\gamma = 4 \ \mbox{meV}$
respectively. The band gap for
GaN at 300 K is 3.43 eV and the band gap for In$_{0.15}$Ga$_{0.85}$N is 2.87 eV.
We chose the photon energy of the pump laser to lie above the band gap of the
In$_{0.15}$Ga$_{0.85}$N wells but below that of the GaN barriers in order to
preferentially photoexcite carriers in the wells.

The pump laser creates electron hole pairs in the
quantum wells which then rapidly cool through confined optical phonon scattering
and slowly recombine across the band gap. In our simulations we solve the
set of Boltzmann equations in Eq.~(\ref{2D Boltzmann equation}) for the time-dependent
conduction and valence electron distribution functions $f^c_{n}(k)$ and $f^v_{n}(k)$.
The phenomenological electron-hole recombination time is take to be
$\tau_0 = 100 \ \mbox{ps}$ since, in an earlier experimental study,\cite{Sanders05.245302}
electrons and holes in an In$_x$Mn$_{1-x}$As/GaSb heterostructure were found to
recombine at midgap defects on a time scale of $\tau_0 \approx 200 \ \mbox{ps}$.
The time-dependence of the
dielectric functions $\varepsilon_1(\hbar\omega)$ and $\varepsilon_2(\hbar\omega)$
is determined by the time-dependence of the conduction and valence band distribution
functions, $f^c_{n}(k)$ and $f^v_{n}(k)$.

\begin{figure} [tbp]
\includegraphics[scale=.8]{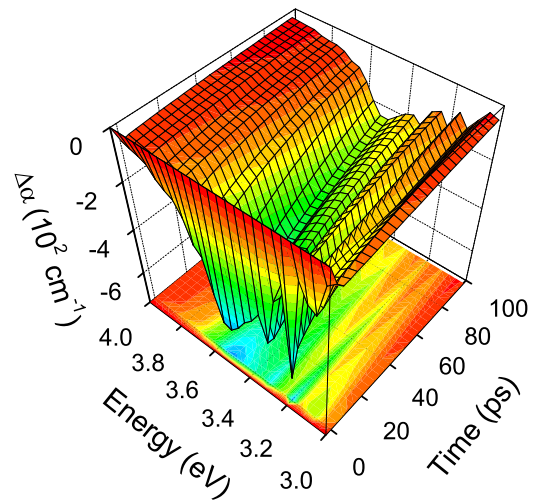}
\caption{(Color Online)
Time dependent differential absorption as a function of delay time and probe
photon energy in a simulated pump-probe experiment in a InGaN/GaN MQW \textit{pin}
diode structure with an external bias of $V \approx 0 \ \textrm{Volts}$.
}
\label{differential absorption coefficient figure}
\end{figure}

The differential absorption coefficient, defined as the change in the absorption
coefficient with respect to its initial thermal equilibrium value, is shown in
Fig.~\ref{differential absorption coefficient figure} as a function of delay
time and probe photon energy. There is a sharp dip in the differential
absorption at the pump photon energy (3.2 eV) which can be attributed to
\emph{bleaching} of the direct transition between photogenerated pairs of electrons
and holes. In the quantum well structure with strong built-in piezoelectric
fields, a given quantum well state is optically coupled to quantum well states in
adjacent wells and in the GaN continuum. When the pump creates electron-hole
pairs, these interband transitions are also blocked giving rise to the negative
differential absorption seen at higher photon energies in
Fig.~\ref{differential absorption coefficient figure}. The relaxation of the
differential absorption as a function of delay time occurs on a time scale
of hundreds of picoseconds and is controlled by intersubband recombination of
the photogenerated electrons and holes.

\subsection{Coherent phonon generation and propagation}

In our simulated pump-probe experiment, electrons and holes are photoexcited
in the conduction and valence bands by the pump laser. These carriers in turn
generate coherent acoustic phonons.
From the photoexcited electron and hole distribution functions $\rho_e(z,t)$
and $\rho_h(z,t)$ defined in Eq.~(\ref{Photogenerated carrier densities}),
we obtain the driving function $S(z,t)$ from Eq.~(\ref{Snu2})
and solve the coherent phonon loaded string equation numerically on
a finite difference space-time grid.

\begin{figure} [tbp]
\includegraphics[scale=.75]{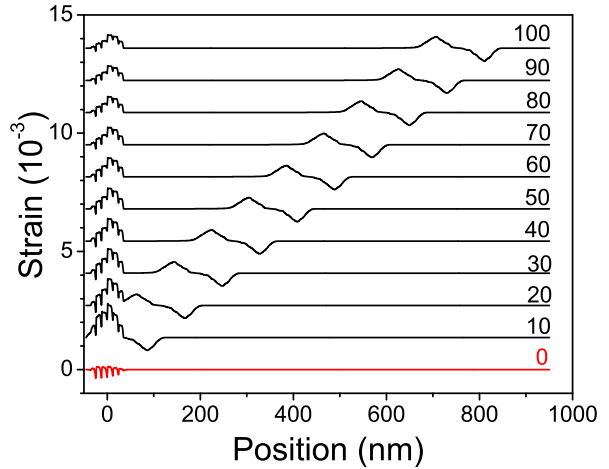}
\caption{(Color Online)
The strain tensor component as a function of position and delay time.
}
\label{propagating strain figure}
\end{figure}

The coherent phonon lattice displacement, $U(z,t)$, obtained from
the loaded string equation (\ref{Loaded string equation}) gives rise to
a propagating strain field. The  coherent phonon strain tensor
component $\varepsilon_{zz}(z,t)$ is shown in
Fig.~\ref{propagating strain figure} as a function of position in the sample
for several equally spaced delay times ranging from 0 to 100 ps.

Following photogeneration of carriers by the pump, a localized strain
appears in the multi-quantum well region as can be seen in
Fig.~\ref{propagating strain figure}. This is due to near steady-state
loading by the driving function at long times. Assuming the driving
function, $S(z,t)$, is approximately  time independent at long times,
the loaded string equation (\ref{Loaded string equation}) can be
integrated once in the steady-state limit. The resulting steady-state
strain is
\begin{equation}
\varepsilon_{zz}(z)=\frac{\partial U(z)}{\partial z}=
-\int_{-\infty}^{z} dz' \ \frac{S(z')}{C_s^2(z')}.
\label{Steady state strain}
\end{equation}
where $C_s(z')$ is the longitudinal acoustic sound speed in the
InGaN quantum wells and $S(z')$ is the approximately time independent driving
function left behind in the multi-quantum well at long times.
The fact that the steady-state strain is localized in the
InGaN/GaN multi-quantum wells follows directly from the sum rule (\ref{Sum rule}).

In addition to the localized strain in the InGaN/GaN multi-quantum well
region of the diode, transient strain pulses are seen to radiate
into the GaN layers at the longitudinal acoustic sound speed. Two
transient strain pulses are generated, one propagating to the
left and the other to the right. The leftward propagating pulse is totally
reflected off the semiconductor-air interface in the p-GaN layer and trails
the rightward propagating pulse as it propagates into the n-GaN layer and the
GaN substrate.

\subsection{Differential reflectivity}

\begin{figure} [tbp]
\includegraphics[scale=.9]{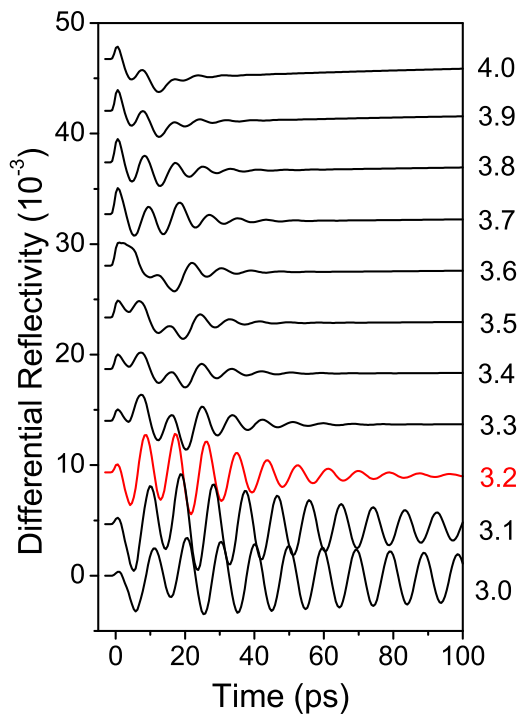}
\caption{(Color Online)
Computed time resolved differential reflectivity as a function of probe delay
for simulated pump-probe experiments on the InGaN/GaN pin diode shown
in Fig.~\ref{MQW diode figure}. The pump photon energy is 3.2 eV
and differential reflectivity is computed for probe-photon energies ranging
from 3.0 to 4.0 eV. The curves are offset for clarity.
}
\label{differential reflectivity figure}
\end{figure}

The computed time resolved differential reflectivity in our simulation
as a function of probe delay is shown in Fig.~\ref{differential reflectivity figure}
for probe energies ranging from 3.0 to 4.0 eV in increments of 0.1 eV.

The oscillation observed in the differential reflectivity can be
attributed to propagation of the strain pulse through the diode.
%
%
The propagating strain tensor shown in Fig.~\ref{propagating strain figure}
alters the local dielectric function as it propagates. The change in the
complex dielectric function is given by
\begin{equation}
\Delta \varepsilon(\hbar\omega,z,t)=
\frac{d \varepsilon(\hbar\omega)}{d \varepsilon_{zz}}
\ \varepsilon_{zz}(z,t),
\label{Dielectric function derivatives wrt strain}
\end{equation}

\begin{figure} [tbp]
\includegraphics[scale=.75]{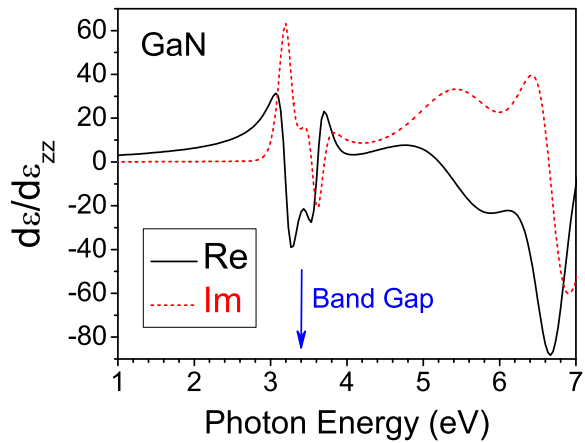}
\caption{(Color Online)
Derivative of the complex GaN dielectric function with respect to strain
as a function of thne probe photon energy. The solid line is the real part
and the dashed line is the imaginary part.
}
\label{GaN dielectric function derivatives wrt strain}
\end{figure}

The total derivative of the complex dielectric function with respect to
strain in Eq.~(\ref{Dielectric function derivatives wrt strain}) measures
the sensitivity of the local dielectric function to small changes in the
strain induced by the coherent phonons
and is obtained from the GaN model dielectric function by
differentiating with respect to $\varepsilon_{zz}$ taking care to eliminate
$\varepsilon_{xx}$ and $\varepsilon_{yy}$ in favor of $\varepsilon_{zz}$. The
real and imaginary parts of $d \varepsilon /d \varepsilon_{zz}$
in GaN are plotted as a function of the probe photon energy in
Fig.~\ref{GaN dielectric function derivatives wrt strain}. The sensitivity
of the dielectric function to changes in the local strain depends
strongly on the probe photon energy and is especially large near the GaN
band gap.
%

The differential reflectivity curves in Fig.~\ref{differential reflectivity figure}
are obtained by solving Maxwell's equations in the structure for the dielectric
function in Eq.~(\ref{Total dielectric function}). We note that the differential
reflectivity signal in Fig.~\ref{differential reflectivity figure} is dominated
by strong coherent phonon oscillations.
In an earlier experimental and theoretical study of time resolved reflectivity
in In$_x$Mn$_{1-x}$As/GaSb heterostructures
(Ref.~\onlinecite{Sanders05.245302}) we found similar oscillations in the
differential reflectivity signal which were a weak perturbation on the background
signal. The background signal in the earlier study was attributed to: (1) enhanced
Drude absorption resulting from the increase in carriers and (2) the relaxation
dynamics associated with the relaxation of highly nonequilibrium
photoexcited carriers. In the time resolved differential reflectivity studies of
two-color pump-probe experiments in InMnAs/GaSb heterostructures described in
Ref.~\onlinecite{Sanders05.245302}, coherent acoustic phonons were generated
\emph{exclusively} through the deformation potential electron phonon
interaction.\cite{Sanders05.245302} The strong differential reflectivity
oscillations seen in the present study, in contrast to the earlier one, can
be attributed to an \emph{order of magnitude enhancement} in the coherent
phonon driving function, $S(z,t)$, due to the strong piezoelectric electron-phonon
interaction term in Eq.~(\ref{Snu2}).

The reflectivity oscillations can be qualitatively understood as
follows. The propagating strain pulse in Fig.~\ref{propagating strain figure}
gives rise to a perturbation in the GaN dielectric function which
propagates at the acoustic sound speed. Our diode structure with the propagating
perturbation in the dielectric function acts as a
Fabry-Perot interferometer and the period of the reflectivity oscillations
due to the propagating coherent acoustic phonon wavepacket is
approximately \cite{Yahng02.4723}
\begin{equation}
T = \frac{\lambda}{2 \ C_s \ n(\lambda)}
\label{Oscillation period}
\end{equation}
where $\lambda=2\pi \ c/\omega$ is the probe wavelength,
$\hbar\omega$ is the probe photon energy,
$C_s$ is the LA sound speed in GaN and $n(\lambda)$ is the wavelength
dependent refractive index. The refractive index can be obtained from
the GaN model dielectric function in
Fig.~\ref{GaN dielectric function figure} as \cite{Haug}
\begin{equation}
n(\lambda)=\sqrt{ \frac{1}{2} \left(
\varepsilon_1(\lambda)+\sqrt{
\varepsilon_1(\lambda)^2 + \varepsilon_2(\lambda)^2}
\right) }
\label{Index of refraction}
\end{equation}

\begin{figure} [tbp]
\includegraphics[scale=.9]{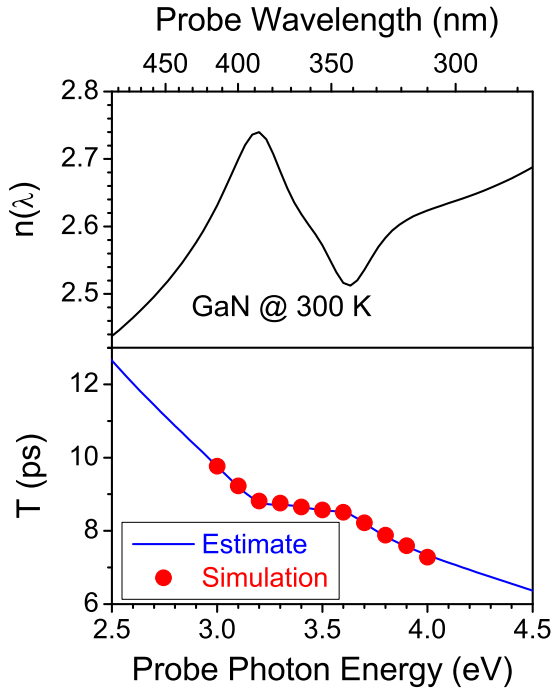}
\caption{(Color Online)
Index of refraction in GaN (upper panel) and period of coherent phonon induced
reflectivity oscillations (lower panel) as functions of
probe photon energy. The solid curve in the lower panel
is based on Eq.~(\ref{Oscillation period}) while the dots are obtained
from the simulations described in the text and presented in
Fig.~\ref{differential reflectivity figure}.
}
\label{reflectivity period figure}
\end{figure}

The index of refraction in GaN obtained from the model dielectric function
in Fig.~\ref{GaN dielectric function figure} and using
Eq.~(\ref{Index of refraction}) is shown in the upper panel of
Fig.~\ref{reflectivity period figure} as a function of probe photon energy
and the period of the coherent phonon induced reflectivity oscillations based
on the estimate of Eq.~(\ref{Oscillation period}) is shown as a solid line
in the lower panel. The period of the reflectivity oscillations obtained from
our simulations and presented in Fig.~\ref{differential reflectivity figure}
are plotted in the lower panel of Fig.~\ref{reflectivity period figure} as
dots. As we can see, the simulation results agree very closely with the
simple estimate in Eq.~(\ref{Oscillation period}).

In Fig.~\ref{differential reflectivity figure}, the coherent phonon induced
reflectivity oscillations are observed to decay as a function of probe delay
and we note that the temporal decay of the reflectivity signal is more rapid
at higher probe photon energies. This has nothing to do with decay of the
propagating coherent phonon strain pulse, but rather can be attributed to
the fact that the absorption coefficient in GaN is rapidly increasing with
probe photon energy above 3.0 eV as can be inferred from the imaginary part
of the GaN model dielectric function in Fig.~\ref{GaN dielectric function figure}.
Hence, at shorter wavelengths, the probe can not penetrate into the sample
to detect the phonons.

\subsection{Coherent control of differential reflectivity}

\begin{figure} [tbp]
\includegraphics[scale=.77]{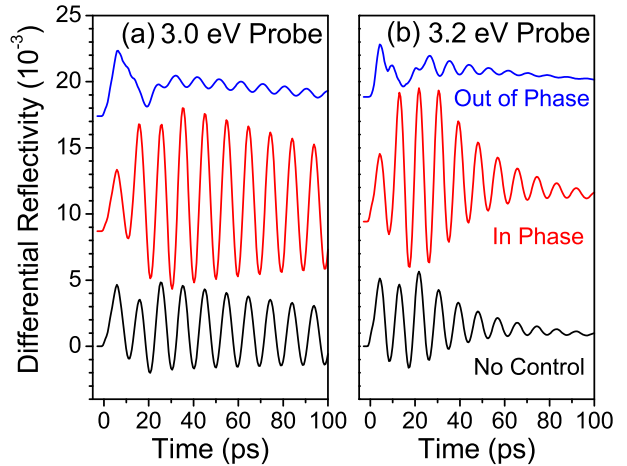}
\caption{(Color Online)
Differential reflectivity as a function of probe delay for probe photon
energies of (a) 3.0 eV and (b) 3.2 eV. The curves are offset for clarity.
The pump photon energies are 3.2 eV in all cases. The lower curves show the
differential reflectivity in the absence of a control pulse while the middle
and upper curves are, respectively, the differential reflectivity with a
3.2 eV control pulse in phase and out of phase with the reflectivity
oscillations at the probe photon energy. The best control results for
destructive interference are achieved with control pulses having approximately
twice the fluence of the initial pump pulse.
}
\label{reflectivity control figure}
\end{figure}

Coherent optical control of coherent acoustic phonon induced differential transmission
oscillations by means of femtosecond laser sources has been experimentally demonstrated
in InGaN/GaN multi-quantum well structures in Refs.~\onlinecite{Sun01.1201} and
\onlinecite{Ozgur01.5604}. In Ref.~\onlinecite{Sun01.1201} control of
coherent differential transmission oscillations is achieved in
a 14 period InGaN/GaN multi-quantum well sample by means of a control
pulse whose time delay and intensity could be controlled independently.
\cite{Sun01.1201,Chern04.339}

In this paper, we investigate the feasibility of optical control of the
differential reflectivity oscillations in the \textit{pin} diode structure shown in\
Fig.~\ref{MQW diode figure}. We consider a simple control scenario in which a
3.2 eV pump pulse is followed by a delayed 3.2 eV control pulse with the same
Gaussian lineshape as the pump but with possibly a different fluence.

Figs.~\ref{reflectivity control figure} (a) and (b) illustrate how the simulated
time resolved differential reflectivity signals seen in
Fig.~\ref{differential reflectivity figure} can be controlled by means
of a delayed control pulse for probe photon energies of 3.0 and 3.2 eV,
respectively.
The curves at the bottom of the figures are the time resolved
differential reflectivity signals in the absence of a control pulse and are
the same as the corresponding curves in Fig.~\ref{differential reflectivity figure}.
The upper two curves in Figs.~\ref{reflectivity control figure} (a) and (b) are
simulated transient differential reflectivity signals in a pump-probe experiment
with a delayed control pump in phase and out of phase with the reflectivity
oscillation period. Both initial and control pulses are 1 ps Gaussian pulses
with linewidths of 4 meV and a photon energy of 3.2 eV. We varied the fluence
of the control pulse relative to the initial pulse in order to maximize destructive
interference in the differential reflectivity for the out of phase cases. We
got the best results using a control pump having twice the fluence of the initial
pump. The middle two curves in the Figs.~\ref{reflectivity control figure} (a) and (b)
show the time resolved differential reflectivity when the control pulses are
in phase with the period of the reflectivity oscillations. As we can see the
time resolved differential reflectivity curves in this case are enhanced due
to constructive interference.

\section{Summary and Conclusions}
\label{Conclusions section}

In summary, we have simulated time-dependent two-color differential reflectivity
experiments on a pseudomorphically strained piezoelectric InGaN/GaN multi-quantum
well in a \textit{pin} diode including bothe the background and
oscillating coherent phonon signal. We perform simulations in which the pump laser
creates electron-hole pairs in the strained InGaN wells by pumping below the GaN
band gap. We conclude that very large amplitude differential reflectivity oscillations
result from the generation of coherent acoustic phonon wavepackets in the
multi-quantum well and their subsequent propagation into the diode.
The propagation of these coherent, localized strain pulses into the diode results
in a position- and frequency-dependent dielectric function which propagates
at the LA phonon sound speed. The diode with the propagating disturbance acts
like a Fabry-Perot interferometer and the period of the differential reflectivity
oscillations, determined from a simple geometrical optics argument, is proportional
to the probe wavelength.

To take into account the time dependent background differential
reflectivity, we modeled the two-color pump-probe reflectivity
experiments in a Boltzmann equation formalism.  Electronic structure
in the InGaN/GaN multi-quantum well was calculated using $\vec k \cdot \vec p$
theory. We included 1) photogeneration of hot carriers in the
multi-quantum wells by a pump laser and 2) their subsequent cooling
by emission and absorption of confined LO phonons. Nonradiative recombination
of electron-hole pairs was also included in a simple relaxation time
approximation. We found that the background differential reflectivity signal
in our simulated diode for probe photon energies near the band gap was smaller
than the amplitude of the coherent phonon induced differential reflectivity
oscillations.

We demonstrate that coherent optical control of the time resolved differential
reflectivity can be achieved. We have investigated the feasibility of optically
controlling the differential reflectivity signal using a delayed
control pulse with the same photon energy and linewidth as the pump but with
a different fluence. By applying the control pulse with a time delay in phase
and out of phase with the differential reflectivity oscillations, the differential
reflectivity signal can be selectively amplified or suppressed.


\begin{acknowledgments}
This work was supported by NSF through DMR-0325474.
\end{acknowledgments}

\bibliography{paper}

\begin{thebibliography}{44}
\expandafter\ifx\csname natexlab\endcsname\relax\def\natexlab#1{#1}\fi
\expandafter\ifx\csname bibnamefont\endcsname\relax
  \def\bibnamefont#1{#1}\fi
\expandafter\ifx\csname bibfnamefont\endcsname\relax
  \def\bibfnamefont#1{#1}\fi
\expandafter\ifx\csname citenamefont\endcsname\relax
  \def\citenamefont#1{#1}\fi
\expandafter\ifx\csname url\endcsname\relax
  \def\url#1{\texttt{#1}}\fi
\expandafter\ifx\csname urlprefix\endcsname\relax\def\urlprefix{URL }\fi
\providecommand{\bibinfo}[2]{#2}
\providecommand{\eprint}[2][]{\url{#2}}

\bibitem[{\citenamefont{Dekorsy et~al.}(1993)\citenamefont{Dekorsy, Pfeifer,
  Kutt, and Kurz}}]{Dekorsy93.3842}
\bibinfo{author}{\bibfnamefont{T.}~\bibnamefont{Dekorsy}},
  \bibinfo{author}{\bibfnamefont{T.}~\bibnamefont{Pfeifer}},
  \bibinfo{author}{\bibfnamefont{W.}~\bibnamefont{Kutt}}, \bibnamefont{and}
  \bibinfo{author}{\bibfnamefont{H.}~\bibnamefont{Kurz}},
  \bibinfo{journal}{Phys. Rev. B} \textbf{\bibinfo{volume}{47}},
  \bibinfo{pages}{3842} (\bibinfo{year}{1993}).

\bibitem[{\citenamefont{Kuznetsov and Stanton}(1995)}]{Kuznetsov95.7555}
\bibinfo{author}{\bibfnamefont{A.~V.} \bibnamefont{Kuznetsov}}
  \bibnamefont{and} \bibinfo{author}{\bibfnamefont{C.~J.}
  \bibnamefont{Stanton}}, \bibinfo{journal}{Phys. Rev. B}
  \textbf{\bibinfo{volume}{51}}, \bibinfo{pages}{7555} (\bibinfo{year}{1995}).

\bibitem[{\citenamefont{Chern et~al.}(2004)\citenamefont{Chern, Sun, Sanders,
  and Stanton}}]{Chern04.339}
\bibinfo{author}{\bibfnamefont{G.-W.} \bibnamefont{Chern}},
  \bibinfo{author}{\bibfnamefont{C.-K.} \bibnamefont{Sun}},
  \bibinfo{author}{\bibfnamefont{G.~D.} \bibnamefont{Sanders}},
  \bibnamefont{and} \bibinfo{author}{\bibfnamefont{C.~J.}
  \bibnamefont{Stanton}}, in \emph{\bibinfo{booktitle}{Topics in Applied
  Physics}}, edited by \bibinfo{editor}{\bibfnamefont{K.-T.}
  \bibnamefont{Tsen}} (\bibinfo{publisher}{Springer-Verlag},
  \bibinfo{address}{New York}, \bibinfo{year}{2004}),
  vol.~\bibinfo{volume}{92}, pp. \bibinfo{pages}{339--394}.

\bibitem[{\citenamefont{Stanton et~al.}(2003)\citenamefont{Stanton, Sanders,
  Liu, Chern, Sun, Yahng, Jho, Sohn, Oh, and Kim}}]{Stanton03.525}
\bibinfo{author}{\bibfnamefont{C.~J.} \bibnamefont{Stanton}},
  \bibinfo{author}{\bibfnamefont{G.~D.} \bibnamefont{Sanders}},
  \bibinfo{author}{\bibfnamefont{R.}~\bibnamefont{Liu}},
  \bibinfo{author}{\bibfnamefont{G.-W.} \bibnamefont{Chern}},
  \bibinfo{author}{\bibfnamefont{C.-K.} \bibnamefont{Sun}},
  \bibinfo{author}{\bibfnamefont{J.~S.} \bibnamefont{Yahng}},
  \bibinfo{author}{\bibfnamefont{Y.~D.} \bibnamefont{Jho}},
  \bibinfo{author}{\bibfnamefont{J.~Y.} \bibnamefont{Sohn}},
  \bibinfo{author}{\bibfnamefont{E.}~\bibnamefont{Oh}}, \bibnamefont{and}
  \bibinfo{author}{\bibfnamefont{D.~S.} \bibnamefont{Kim}},
  \bibinfo{journal}{Superlatt. and Microstruct.} \textbf{\bibinfo{volume}{34}},
  \bibinfo{pages}{525} (\bibinfo{year}{2003}).

\bibitem[{\citenamefont{Yahng et~al.}(2002)\citenamefont{Yahng, Jho, Yee, Oh,
  Woo, Kim, Sanders, and Stanton}}]{Yahng02.4723}
\bibinfo{author}{\bibfnamefont{J.~S.} \bibnamefont{Yahng}},
  \bibinfo{author}{\bibfnamefont{Y.~D.} \bibnamefont{Jho}},
  \bibinfo{author}{\bibfnamefont{K.~J.} \bibnamefont{Yee}},
  \bibinfo{author}{\bibfnamefont{E.}~\bibnamefont{Oh}},
  \bibinfo{author}{\bibfnamefont{J.~C.} \bibnamefont{Woo}},
  \bibinfo{author}{\bibfnamefont{D.~S.} \bibnamefont{Kim}},
  \bibinfo{author}{\bibfnamefont{G.~D.} \bibnamefont{Sanders}},
  \bibnamefont{and} \bibinfo{author}{\bibfnamefont{C.~J.}
  \bibnamefont{Stanton}}, \bibinfo{journal}{Appl. Phys. Lett.}
  \textbf{\bibinfo{volume}{80}}, \bibinfo{pages}{4723} (\bibinfo{year}{2002}).

\bibitem[{\citenamefont{Liu et~al.}(2005)\citenamefont{Liu, Kim, Sanders,
  Stanton, Yahng, Jho, Yee, Oh, and Kim}}]{Liu05.195335}
\bibinfo{author}{\bibfnamefont{R.}~\bibnamefont{Liu}},
  \bibinfo{author}{\bibfnamefont{C.~S.} \bibnamefont{Kim}},
  \bibinfo{author}{\bibfnamefont{G.~D.} \bibnamefont{Sanders}},
  \bibinfo{author}{\bibfnamefont{C.~J.} \bibnamefont{Stanton}},
  \bibinfo{author}{\bibfnamefont{J.~S.} \bibnamefont{Yahng}},
  \bibinfo{author}{\bibfnamefont{Y.~D.} \bibnamefont{Jho}},
  \bibinfo{author}{\bibfnamefont{K.~J.} \bibnamefont{Yee}},
  \bibinfo{author}{\bibfnamefont{E.}~\bibnamefont{Oh}}, \bibnamefont{and}
  \bibinfo{author}{\bibfnamefont{D.~S.} \bibnamefont{Kim}},
  \bibinfo{journal}{Phys. Rev. B} \textbf{\bibinfo{volume}{72}},
  \bibinfo{pages}{195335} (\bibinfo{year}{2005}).

\bibitem[{\citenamefont{Sun et~al.}(2001)\citenamefont{Sun, Huang, Liang,
  Abare, and DenBaars}}]{Sun01.1201}
\bibinfo{author}{\bibfnamefont{C.-K.} \bibnamefont{Sun}},
  \bibinfo{author}{\bibfnamefont{Y.-K.} \bibnamefont{Huang}},
  \bibinfo{author}{\bibfnamefont{J.-C.} \bibnamefont{Liang}},
  \bibinfo{author}{\bibfnamefont{A.}~\bibnamefont{Abare}}, \bibnamefont{and}
  \bibinfo{author}{\bibfnamefont{S.~P.} \bibnamefont{DenBaars}},
  \bibinfo{journal}{Appl. Phys. Lett.} \textbf{\bibinfo{volume}{78}},
  \bibinfo{pages}{1201} (\bibinfo{year}{2001}).

\bibitem[{\citenamefont{\"{U}mit \"{O}zg\"{u}r
  et~al.}(2001)\citenamefont{\"{U}mit \"{O}zg\"{u}r, Lee, and
  Everitt}}]{Ozgur01.5604}
\bibinfo{author}{\bibnamefont{\"{U}mit \"{O}zg\"{u}r}},
  \bibinfo{author}{\bibfnamefont{C.-W.} \bibnamefont{Lee}}, \bibnamefont{and}
  \bibinfo{author}{\bibfnamefont{H.~O.} \bibnamefont{Everitt}},
  \bibinfo{journal}{Phys. Rev. Lett.} \textbf{\bibinfo{volume}{86}},
  \bibinfo{pages}{5604} (\bibinfo{year}{2001}).

\bibitem[{\citenamefont{Sanders et~al.}(2005)\citenamefont{Sanders, Stanton,
  Wang, Kono, Oiwa, and Munekata}}]{Sanders05.245302}
\bibinfo{author}{\bibfnamefont{G.~D.} \bibnamefont{Sanders}},
  \bibinfo{author}{\bibfnamefont{C.~J.} \bibnamefont{Stanton}},
  \bibinfo{author}{\bibfnamefont{J.}~\bibnamefont{Wang}},
  \bibinfo{author}{\bibfnamefont{J.}~\bibnamefont{Kono}},
  \bibinfo{author}{\bibfnamefont{A.}~\bibnamefont{Oiwa}}, \bibnamefont{and}
  \bibinfo{author}{\bibfnamefont{H.}~\bibnamefont{Munekata}},
  \bibinfo{journal}{Phys. Rev. B} \textbf{\bibinfo{volume}{72}},
  \bibinfo{pages}{245302} (\bibinfo{year}{2005}).

\bibitem[{\citenamefont{Wang et~al.}(2005)\citenamefont{Wang, Hashimoto, Kono,
  Oiwa, Munekata, Sanders, and Stanton}}]{Wang05.153311}
\bibinfo{author}{\bibfnamefont{J.}~\bibnamefont{Wang}},
  \bibinfo{author}{\bibfnamefont{Y.}~\bibnamefont{Hashimoto}},
  \bibinfo{author}{\bibfnamefont{J.}~\bibnamefont{Kono}},
  \bibinfo{author}{\bibfnamefont{A.}~\bibnamefont{Oiwa}},
  \bibinfo{author}{\bibfnamefont{H.}~\bibnamefont{Munekata}},
  \bibinfo{author}{\bibfnamefont{G.~D.} \bibnamefont{Sanders}},
  \bibnamefont{and} \bibinfo{author}{\bibfnamefont{C.~J.}
  \bibnamefont{Stanton}}, \bibinfo{journal}{Phys. Rev. B}
  \textbf{\bibinfo{volume}{72}}, \bibinfo{pages}{153311}
  (\bibinfo{year}{2005}).

\bibitem[{\citenamefont{Sanders et~al.}(2001)\citenamefont{Sanders, Stanton,
  and Kim}}]{Sanders01.235316}
\bibinfo{author}{\bibfnamefont{G.~D.} \bibnamefont{Sanders}},
  \bibinfo{author}{\bibfnamefont{C.~J.} \bibnamefont{Stanton}},
  \bibnamefont{and} \bibinfo{author}{\bibfnamefont{C.~S.} \bibnamefont{Kim}},
  \bibinfo{journal}{Phys. Rev. B} \textbf{\bibinfo{volume}{64}},
  \bibinfo{pages}{235316} (\bibinfo{year}{2001}).

\bibitem[{\citenamefont{Sanders et~al.}(2002)\citenamefont{Sanders, Stanton,
  and Kim}}]{Sanders02.079903}
\bibinfo{author}{\bibfnamefont{G.~D.} \bibnamefont{Sanders}},
  \bibinfo{author}{\bibfnamefont{C.~J.} \bibnamefont{Stanton}},
  \bibnamefont{and} \bibinfo{author}{\bibfnamefont{C.~S.} \bibnamefont{Kim}},
  \bibinfo{journal}{Phys. Rev. B} \textbf{\bibinfo{volume}{66}},
  \bibinfo{pages}{079903} (\bibinfo{year}{2002}).

\bibitem[{\citenamefont{Jho et~al.}(2001)\citenamefont{Jho, Yahng, Oh, and
  Kim}}]{Jho01.1130}
\bibinfo{author}{\bibfnamefont{Y.~D.} \bibnamefont{Jho}},
  \bibinfo{author}{\bibfnamefont{J.~S.} \bibnamefont{Yahng}},
  \bibinfo{author}{\bibfnamefont{E.}~\bibnamefont{Oh}}, \bibnamefont{and}
  \bibinfo{author}{\bibfnamefont{D.~S.} \bibnamefont{Kim}},
  \bibinfo{journal}{Appl. Phys. Lett.} \textbf{\bibinfo{volume}{79}},
  \bibinfo{pages}{1130} (\bibinfo{year}{2001}).

\bibitem[{\citenamefont{Jho et~al.}(2002)\citenamefont{Jho, Yahng, Oh, and
  Kim}}]{Jho02.035334}
\bibinfo{author}{\bibfnamefont{Y.~D.} \bibnamefont{Jho}},
  \bibinfo{author}{\bibfnamefont{J.~S.} \bibnamefont{Yahng}},
  \bibinfo{author}{\bibfnamefont{E.}~\bibnamefont{Oh}}, \bibnamefont{and}
  \bibinfo{author}{\bibfnamefont{D.~S.} \bibnamefont{Kim}},
  \bibinfo{journal}{Phys. Rev. B} \textbf{\bibinfo{volume}{66}},
  \bibinfo{pages}{035334} (\bibinfo{year}{2002}).

\bibitem[{\citenamefont{Jeon et~al.}(1997)\citenamefont{Jeon, Lee, Sirenko,
  Kim, and Littlejohn}}]{Jeon97.386}
\bibinfo{author}{\bibfnamefont{J.~B.} \bibnamefont{Jeon}},
  \bibinfo{author}{\bibfnamefont{B.~C.} \bibnamefont{Lee}},
  \bibinfo{author}{\bibfnamefont{M.}~\bibnamefont{Sirenko}},
  \bibinfo{author}{\bibfnamefont{K.~W.} \bibnamefont{Kim}}, \bibnamefont{and}
  \bibinfo{author}{\bibfnamefont{M.~A.} \bibnamefont{Littlejohn}},
  \bibinfo{journal}{J. Appl. Phys.} \textbf{\bibinfo{volume}{82}},
  \bibinfo{pages}{386} (\bibinfo{year}{1997}).

\bibitem[{\citenamefont{Chuang and Chang}(1996{\natexlab{a}})}]{Chuang96.2491}
\bibinfo{author}{\bibfnamefont{S.~L.} \bibnamefont{Chuang}} \bibnamefont{and}
  \bibinfo{author}{\bibfnamefont{C.~S.} \bibnamefont{Chang}},
  \bibinfo{journal}{Phys. Rev. B} \textbf{\bibinfo{volume}{54}},
  \bibinfo{pages}{2491} (\bibinfo{year}{1996}{\natexlab{a}}).

\bibitem[{\citenamefont{Chuang and Chang}(1996{\natexlab{b}})}]{Chuang96.1657}
\bibinfo{author}{\bibfnamefont{S.~L.} \bibnamefont{Chuang}} \bibnamefont{and}
  \bibinfo{author}{\bibfnamefont{C.~S.} \bibnamefont{Chang}},
  \bibinfo{journal}{Appl. Phys. Lett.} \textbf{\bibinfo{volume}{68}},
  \bibinfo{pages}{1657} (\bibinfo{year}{1996}{\natexlab{b}}).

\bibitem[{\citenamefont{Vurgaftman et~al.}(2001)\citenamefont{Vurgaftman,
  Meyer, and Ram-Mohan}}]{Vurgaftman01.5815}
\bibinfo{author}{\bibfnamefont{I.}~\bibnamefont{Vurgaftman}},
  \bibinfo{author}{\bibfnamefont{J.~R.} \bibnamefont{Meyer}}, \bibnamefont{and}
  \bibinfo{author}{\bibfnamefont{L.~R.} \bibnamefont{Ram-Mohan}},
  \bibinfo{journal}{J. Appl. Phys.} \textbf{\bibinfo{volume}{89}},
  \bibinfo{pages}{5815} (\bibinfo{year}{2001}).

\bibitem[{\citenamefont{Doshi et~al.}(1998)\citenamefont{Doshi, Brennan,
  Bicknell-Tassius, and Grunthaner}}]{Doshi98.2784}
\bibinfo{author}{\bibfnamefont{B.}~\bibnamefont{Doshi}},
  \bibinfo{author}{\bibfnamefont{K.~F.} \bibnamefont{Brennan}},
  \bibinfo{author}{\bibfnamefont{R.}~\bibnamefont{Bicknell-Tassius}},
  \bibnamefont{and}
  \bibinfo{author}{\bibfnamefont{F.}~\bibnamefont{Grunthaner}},
  \bibinfo{journal}{Appl. Phys. Lett.} \textbf{\bibinfo{volume}{73}},
  \bibinfo{pages}{2784} (\bibinfo{year}{1998}).

\bibitem[{\citenamefont{Martin et~al.}(1996)\citenamefont{Martin, Botchkarev,
  Rockett, and Morko\c{c}}}]{Martin96.2541}
\bibinfo{author}{\bibfnamefont{G.}~\bibnamefont{Martin}},
  \bibinfo{author}{\bibfnamefont{A.}~\bibnamefont{Botchkarev}},
  \bibinfo{author}{\bibfnamefont{A.}~\bibnamefont{Rockett}}, \bibnamefont{and}
  \bibinfo{author}{\bibfnamefont{H.}~\bibnamefont{Morko\c{c}}},
  \bibinfo{journal}{Appl. Phys. Lett.} \textbf{\bibinfo{volume}{68}},
  \bibinfo{pages}{2541} (\bibinfo{year}{1996}).

\bibitem[{\citenamefont{Bernardini et~al.}(1997)\citenamefont{Bernardini,
  Fiorentini, and Vanderbilt}}]{Bernardini97.3958}
\bibinfo{author}{\bibfnamefont{F.}~\bibnamefont{Bernardini}},
  \bibinfo{author}{\bibfnamefont{V.}~\bibnamefont{Fiorentini}},
  \bibnamefont{and}
  \bibinfo{author}{\bibfnamefont{D.}~\bibnamefont{Vanderbilt}},
  \bibinfo{journal}{Phys. Rev. Lett.} \textbf{\bibinfo{volume}{79}},
  \bibinfo{pages}{3958} (\bibinfo{year}{1997}).

\bibitem[{\citenamefont{Yu et~al.}(1997)\citenamefont{Yu, Wang, Ishikawa,
  Umeno, Soga, Egawa, Watanabe, and Jimbo}}]{Yu97.3209}
\bibinfo{author}{\bibfnamefont{G.}~\bibnamefont{Yu}},
  \bibinfo{author}{\bibfnamefont{G.}~\bibnamefont{Wang}},
  \bibinfo{author}{\bibfnamefont{H.}~\bibnamefont{Ishikawa}},
  \bibinfo{author}{\bibfnamefont{M.}~\bibnamefont{Umeno}},
  \bibinfo{author}{\bibfnamefont{T.}~\bibnamefont{Soga}},
  \bibinfo{author}{\bibfnamefont{T.}~\bibnamefont{Egawa}},
  \bibinfo{author}{\bibfnamefont{J.}~\bibnamefont{Watanabe}}, \bibnamefont{and}
  \bibinfo{author}{\bibfnamefont{T.}~\bibnamefont{Jimbo}},
  \bibinfo{journal}{Appl. Phys. Lett.} \textbf{\bibinfo{volume}{70}},
  \bibinfo{pages}{3209} (\bibinfo{year}{1997}).

\bibitem[{\citenamefont{Yang et~al.}(2002)\citenamefont{Yang, Shen, Qian, Pang,
  Ogawa, and Guo}}]{Yang02.9803}
\bibinfo{author}{\bibfnamefont{H.~F.} \bibnamefont{Yang}},
  \bibinfo{author}{\bibfnamefont{W.~Z.} \bibnamefont{Shen}},
  \bibinfo{author}{\bibfnamefont{Z.~G.} \bibnamefont{Qian}},
  \bibinfo{author}{\bibfnamefont{Q.~J.} \bibnamefont{Pang}},
  \bibinfo{author}{\bibfnamefont{H.}~\bibnamefont{Ogawa}}, \bibnamefont{and}
  \bibinfo{author}{\bibfnamefont{Q.~X.} \bibnamefont{Guo}},
  \bibinfo{journal}{J. Appl. Phys.} \textbf{\bibinfo{volume}{91}},
  \bibinfo{pages}{9803} (\bibinfo{year}{2002}).

\bibitem[{\citenamefont{Varshni}(1967)}]{Varshni67.149}
\bibinfo{author}{\bibfnamefont{Y.~P.} \bibnamefont{Varshni}},
  \bibinfo{journal}{Physica} \textbf{\bibinfo{volume}{34}},
  \bibinfo{pages}{149} (\bibinfo{year}{1967}).

\bibitem[{\citenamefont{Walukiewicz}(2004)}]{Walukiewicz04.300}
\bibinfo{author}{\bibfnamefont{W.}~\bibnamefont{Walukiewicz}},
  \bibinfo{journal}{Physica E} \textbf{\bibinfo{volume}{20}},
  \bibinfo{pages}{300} (\bibinfo{year}{2004}).

\bibitem[{\citenamefont{Davydov et~al.}(2002)\citenamefont{Davydov, Klochikhin,
  Seisyan, Emtsev, Inanov, Bechstedt, Furthmuller, Harima, Mudryi, Aderhold
  et~al.}}]{Davydov02.1}
\bibinfo{author}{\bibfnamefont{V.~Y.} \bibnamefont{Davydov}},
  \bibinfo{author}{\bibfnamefont{A.~A.} \bibnamefont{Klochikhin}},
  \bibinfo{author}{\bibfnamefont{R.~P.} \bibnamefont{Seisyan}},
  \bibinfo{author}{\bibfnamefont{V.~V.} \bibnamefont{Emtsev}},
  \bibinfo{author}{\bibfnamefont{S.~V.} \bibnamefont{Inanov}},
  \bibinfo{author}{\bibfnamefont{F.}~\bibnamefont{Bechstedt}},
  \bibinfo{author}{\bibfnamefont{J.}~\bibnamefont{Furthmuller}},
  \bibinfo{author}{\bibfnamefont{H.}~\bibnamefont{Harima}},
  \bibinfo{author}{\bibfnamefont{A.~V.} \bibnamefont{Mudryi}},
  \bibinfo{author}{\bibfnamefont{J.}~\bibnamefont{Aderhold}},
  \bibnamefont{et~al.}, \bibinfo{journal}{Phys. Stat. Solidi. (b)}
  \textbf{\bibinfo{volume}{229}}, \bibinfo{pages}{R1} (\bibinfo{year}{2002}).

\bibitem[{\citenamefont{Wu et~al.}(2002)\citenamefont{Wu, Walukiewicz, Yu, III,
  Haller, Lu, Schaff, Saito, and Nanishi}}]{Wu02.3967}
\bibinfo{author}{\bibfnamefont{J.}~\bibnamefont{Wu}},
  \bibinfo{author}{\bibfnamefont{W.}~\bibnamefont{Walukiewicz}},
  \bibinfo{author}{\bibfnamefont{K.~M.} \bibnamefont{Yu}},
  \bibinfo{author}{\bibfnamefont{J.~W.~A.} \bibnamefont{III}},
  \bibinfo{author}{\bibfnamefont{E.~E.} \bibnamefont{Haller}},
  \bibinfo{author}{\bibfnamefont{H.}~\bibnamefont{Lu}},
  \bibinfo{author}{\bibfnamefont{W.~J.} \bibnamefont{Schaff}},
  \bibinfo{author}{\bibfnamefont{Y.}~\bibnamefont{Saito}}, \bibnamefont{and}
  \bibinfo{author}{\bibfnamefont{Y.}~\bibnamefont{Nanishi}},
  \bibinfo{journal}{Appl. Phys. Lett.} \textbf{\bibinfo{volume}{80}},
  \bibinfo{pages}{3967} (\bibinfo{year}{2002}).

\bibitem[{\citenamefont{Matsuoka et~al.}(2004)\citenamefont{Matsuoka, Okamoto,
  Nakao, Harima, and Kurimoto}}]{Matsuoka02.1246}
\bibinfo{author}{\bibfnamefont{T.}~\bibnamefont{Matsuoka}},
  \bibinfo{author}{\bibfnamefont{H.}~\bibnamefont{Okamoto}},
  \bibinfo{author}{\bibfnamefont{N.}~\bibnamefont{Nakao}},
  \bibinfo{author}{\bibfnamefont{H.}~\bibnamefont{Harima}}, \bibnamefont{and}
  \bibinfo{author}{\bibfnamefont{E.}~\bibnamefont{Kurimoto}},
  \bibinfo{journal}{Appl. Phys. Lett.} \textbf{\bibinfo{volume}{81}},
  \bibinfo{pages}{1246} (\bibinfo{year}{2004}).

\bibitem[{\citenamefont{Nakamura and Fasol}(1997)}]{Nakamura}
\bibinfo{author}{\bibfnamefont{S.}~\bibnamefont{Nakamura}} \bibnamefont{and}
  \bibinfo{author}{\bibfnamefont{G.}~\bibnamefont{Fasol}},
  \emph{\bibinfo{title}{The blue laser diode: GaN based emitters and lasers}}
  (\bibinfo{publisher}{Springer-Verlag}, \bibinfo{address}{Berlin},
  \bibinfo{year}{1997}).

\bibitem[{\citenamefont{Smith and Mailhiot}(1988)}]{Smith88.2717}
\bibinfo{author}{\bibfnamefont{D.~L.} \bibnamefont{Smith}} \bibnamefont{and}
  \bibinfo{author}{\bibfnamefont{C.}~\bibnamefont{Mailhiot}},
  \bibinfo{journal}{J. Appl. Phys.} \textbf{\bibinfo{volume}{63}},
  \bibinfo{pages}{2717} (\bibinfo{year}{1988}).

\bibitem[{\citenamefont{Lefebvre et~al.}(2001)\citenamefont{Lefebvre, Morel,
  Gallart, Taliercio, Allgre, Gil, Mathieu, Damilano, Grandjean, and
  Massies}}]{Lefebvre01.1252}
\bibinfo{author}{\bibfnamefont{P.}~\bibnamefont{Lefebvre}},
  \bibinfo{author}{\bibfnamefont{A.}~\bibnamefont{Morel}},
  \bibinfo{author}{\bibfnamefont{M.}~\bibnamefont{Gallart}},
  \bibinfo{author}{\bibfnamefont{T.}~\bibnamefont{Taliercio}},
  \bibinfo{author}{\bibfnamefont{J.}~\bibnamefont{Allgre}},
  \bibinfo{author}{\bibfnamefont{B.}~\bibnamefont{Gil}},
  \bibinfo{author}{\bibfnamefont{H.}~\bibnamefont{Mathieu}},
  \bibinfo{author}{\bibfnamefont{B.}~\bibnamefont{Damilano}},
  \bibinfo{author}{\bibfnamefont{N.}~\bibnamefont{Grandjean}},
  \bibnamefont{and} \bibinfo{author}{\bibfnamefont{J.}~\bibnamefont{Massies}},
  \bibinfo{journal}{Appl. Phys. Lett.} \textbf{\bibinfo{volume}{78}},
  \bibinfo{pages}{1252} (\bibinfo{year}{2001}).

\bibitem[{\citenamefont{Takeuchi et~al.}(1998)\citenamefont{Takeuchi, Wetzel,
  Yamaguchi, Sakai, Amano, Akasaki, Kaneko, Nakagawa, Yamaoka, and
  Yamada}}]{Takeuchi98.1691}
\bibinfo{author}{\bibfnamefont{T.}~\bibnamefont{Takeuchi}},
  \bibinfo{author}{\bibfnamefont{C.}~\bibnamefont{Wetzel}},
  \bibinfo{author}{\bibfnamefont{S.}~\bibnamefont{Yamaguchi}},
  \bibinfo{author}{\bibfnamefont{H.}~\bibnamefont{Sakai}},
  \bibinfo{author}{\bibfnamefont{H.}~\bibnamefont{Amano}},
  \bibinfo{author}{\bibfnamefont{I.}~\bibnamefont{Akasaki}},
  \bibinfo{author}{\bibfnamefont{Y.}~\bibnamefont{Kaneko}},
  \bibinfo{author}{\bibfnamefont{S.}~\bibnamefont{Nakagawa}},
  \bibinfo{author}{\bibfnamefont{Y.}~\bibnamefont{Yamaoka}}, \bibnamefont{and}
  \bibinfo{author}{\bibfnamefont{N.}~\bibnamefont{Yamada}},
  \bibinfo{journal}{Appl. Phys. Lett.} \textbf{\bibinfo{volume}{73}},
  \bibinfo{pages}{1691} (\bibinfo{year}{1998}).

\bibitem[{\citenamefont{Im et~al.}(1998)\citenamefont{Im, Kollmer, Off, Sohmer,
  Scholz, and Hangleiter}}]{Im98.9435}
\bibinfo{author}{\bibfnamefont{J.~S.} \bibnamefont{Im}},
  \bibinfo{author}{\bibfnamefont{H.}~\bibnamefont{Kollmer}},
  \bibinfo{author}{\bibfnamefont{J.}~\bibnamefont{Off}},
  \bibinfo{author}{\bibfnamefont{A.}~\bibnamefont{Sohmer}},
  \bibinfo{author}{\bibfnamefont{F.}~\bibnamefont{Scholz}}, \bibnamefont{and}
  \bibinfo{author}{\bibfnamefont{A.}~\bibnamefont{Hangleiter}},
  \bibinfo{journal}{Phys. Rev. B} \textbf{\bibinfo{volume}{57}},
  \bibinfo{pages}{R9435} (\bibinfo{year}{1998}).

\bibitem[{\citenamefont{Wetzel et~al.}(2000)\citenamefont{Wetzel, Takeuchi,
  Amano, and Akasaki}}]{Wetzel00.2159}
\bibinfo{author}{\bibfnamefont{C.}~\bibnamefont{Wetzel}},
  \bibinfo{author}{\bibfnamefont{T.}~\bibnamefont{Takeuchi}},
  \bibinfo{author}{\bibfnamefont{H.}~\bibnamefont{Amano}}, \bibnamefont{and}
  \bibinfo{author}{\bibfnamefont{I.}~\bibnamefont{Akasaki}},
  \bibinfo{journal}{Phys. Rev. B} \textbf{\bibinfo{volume}{61}},
  \bibinfo{pages}{2159} (\bibinfo{year}{2000}).

\bibitem[{\citenamefont{Chichibu et~al.}(2000)\citenamefont{Chichibu, Azuhata,
  Sota, Mukai, and Nakamura}}]{Chichibu00.5153}
\bibinfo{author}{\bibfnamefont{S.~F.} \bibnamefont{Chichibu}},
  \bibinfo{author}{\bibfnamefont{T.}~\bibnamefont{Azuhata}},
  \bibinfo{author}{\bibfnamefont{T.}~\bibnamefont{Sota}},
  \bibinfo{author}{\bibfnamefont{T.}~\bibnamefont{Mukai}}, \bibnamefont{and}
  \bibinfo{author}{\bibfnamefont{S.}~\bibnamefont{Nakamura}},
  \bibinfo{journal}{Appl. Phys. Lett.} \textbf{\bibinfo{volume}{88}},
  \bibinfo{pages}{5153} (\bibinfo{year}{2000}).

\bibitem[{\citenamefont{Wright}(1997)}]{Wright97.2833}
\bibinfo{author}{\bibfnamefont{A.~F.} \bibnamefont{Wright}},
  \bibinfo{journal}{J. Appl. Phys.} \textbf{\bibinfo{volume}{82}},
  \bibinfo{pages}{2833} (\bibinfo{year}{1997}).

\bibitem[{\citenamefont{Chuang}(1995)}]{Chuang}
\bibinfo{author}{\bibfnamefont{S.~L.} \bibnamefont{Chuang}},
  \emph{\bibinfo{title}{Physics of Optoelectronic Devices}}
  (\bibinfo{publisher}{Wiley}, \bibinfo{address}{New York},
  \bibinfo{year}{1995}).

\bibitem[{\citenamefont{Sanders and Stanton}(1998)}]{Sanders98.9148}
\bibinfo{author}{\bibfnamefont{G.~D.} \bibnamefont{Sanders}} \bibnamefont{and}
  \bibinfo{author}{\bibfnamefont{C.~J.} \bibnamefont{Stanton}},
  \bibinfo{journal}{Phys. Rev. B} \textbf{\bibinfo{volume}{57}},
  \bibinfo{pages}{9148} (\bibinfo{year}{1998}).

\bibitem[{\citenamefont{Sch\"{a}fer and Wegener}(2002)}]{Schafer}
\bibinfo{author}{\bibfnamefont{W.}~\bibnamefont{Sch\"{a}fer}} \bibnamefont{and}
  \bibinfo{author}{\bibfnamefont{M.}~\bibnamefont{Wegener}},
  \emph{\bibinfo{title}{Semiconductor Optics and Transport Phenomena}}
  (\bibinfo{publisher}{Springer}, \bibinfo{address}{New York},
  \bibinfo{year}{2002}).

\bibitem[{\citenamefont{Djuri\v{s}i\'{c} and Li}(1999)}]{Djurisic99.2848}
\bibinfo{author}{\bibfnamefont{A.~B.} \bibnamefont{Djuri\v{s}i\'{c}}}
  \bibnamefont{and} \bibinfo{author}{\bibfnamefont{E.~H.} \bibnamefont{Li}},
  \bibinfo{journal}{J. Appl. Phys.} \textbf{\bibinfo{volume}{84}},
  \bibinfo{pages}{2848} (\bibinfo{year}{1999}).

\bibitem[{\citenamefont{Thomsen et~al.}(1984)\citenamefont{Thomsen, Strait,
  Vardeny, Maris, Tauc, and Hauser}}]{Thomsen84.989}
\bibinfo{author}{\bibfnamefont{C.}~\bibnamefont{Thomsen}},
  \bibinfo{author}{\bibfnamefont{J.}~\bibnamefont{Strait}},
  \bibinfo{author}{\bibfnamefont{Z.}~\bibnamefont{Vardeny}},
  \bibinfo{author}{\bibfnamefont{H.~J.} \bibnamefont{Maris}},
  \bibinfo{author}{\bibfnamefont{J.}~\bibnamefont{Tauc}}, \bibnamefont{and}
  \bibinfo{author}{\bibfnamefont{J.~J.} \bibnamefont{Hauser}},
  \bibinfo{journal}{Phys. Rev. Lett.} \textbf{\bibinfo{volume}{53}},
  \bibinfo{pages}{989} (\bibinfo{year}{1984}).

\bibitem[{\citenamefont{Thomsen et~al.}(1986)\citenamefont{Thomsen, Grahn,
  Maris, and Tauc}}]{Thomsen86.4129}
\bibinfo{author}{\bibfnamefont{C.}~\bibnamefont{Thomsen}},
  \bibinfo{author}{\bibfnamefont{H.~T.} \bibnamefont{Grahn}},
  \bibinfo{author}{\bibfnamefont{H.~J.} \bibnamefont{Maris}}, \bibnamefont{and}
  \bibinfo{author}{\bibfnamefont{J.}~\bibnamefont{Tauc}},
  \bibinfo{journal}{Phys. Rev. B} \textbf{\bibinfo{volume}{34}},
  \bibinfo{pages}{4129} (\bibinfo{year}{1986}).

\bibitem[{\citenamefont{R\"{o}nnow et~al.}(1999)\citenamefont{R\"{o}nnow,
  Christensen, and Cardona}}]{Ronnow99.5575}
\bibinfo{author}{\bibfnamefont{D.}~\bibnamefont{R\"{o}nnow}},
  \bibinfo{author}{\bibfnamefont{N.~E.} \bibnamefont{Christensen}},
  \bibnamefont{and} \bibinfo{author}{\bibfnamefont{M.}~\bibnamefont{Cardona}},
  \bibinfo{journal}{Phys. Rev. B} \textbf{\bibinfo{volume}{59}},
  \bibinfo{pages}{5575} (\bibinfo{year}{1999}).

\bibitem[{\citenamefont{Haug and Koch}(1993)}]{Haug}
\bibinfo{author}{\bibfnamefont{H.}~\bibnamefont{Haug}} \bibnamefont{and}
  \bibinfo{author}{\bibfnamefont{S.~W.} \bibnamefont{Koch}},
  \emph{\bibinfo{title}{Quantum Theory of the Optical and Electronic Properties
  of Semiconductors}} (\bibinfo{publisher}{World Scientific},
  \bibinfo{address}{New Jersey}, \bibinfo{year}{1993}).

\end{thebibliography}

\end{document}